\documentclass[epj,dvipsnames,nofootinbib,superscriptaddress,doi=false]{svjour}
\usepackage[utf8]{inputenc}
\usepackage[T1]{fontenc}

\usepackage{multicol}

\usepackage{cuted}

\usepackage{tikz}
\usetikzlibrary{shapes,positioning} 
\usetikzlibrary{decorations,decorations.markings}

\usepackage{pgfplots}
\pgfplotsset{compat=newest}

\usepackage{graphicx}
\usepackage{rotating}
\usepackage{pifont}
\usepackage{xcolor}

\usepackage{isotope}

\usepackage{scalefnt}

\usepackage{hyperref}

\usepackage{amsmath}
\usepackage{amssymb}
\usepackage{amsfonts}
\usepackage{bm}
\usepackage{xfrac}

\tikzset{
    >=latex,
    operatornode/.style={draw=black,fill=black,minimum size=7pt,shape=rectangle,inner sep=0pt},
    omeganode/.style={draw=black,fill=black,minimum size=7pt,shape=circle,inner sep=0pt},
    Rnode/.style={draw=black,fill=black,shape=rectangle,inner sep=0pt,text width=1cm},
    Lambdanode/.style={draw=black,fill=black,shape=rectangle,inner sep=0pt,text width=1cm},
    Tnode/.style={draw=black,fill=black,shape=rectangle,inner sep=0pt,text width=0.2cm},
    cross/.style={cross out, very thick,draw=black, minimum size=8*(#1-\pgflinewidth), inner sep=0pt, outer sep=0pt},cross/.default={1pt}
    every node/.append style={font=\small},
    every edge/.append style={very thick},
    arrow/.style={ultra thick, shorten >=5pt,shorten <=5pt,->},
    photon/.style={decorate, decoration={snake}},
    particle/.style={postaction={decorate},
    decoration={markings,mark=at position .56 with {\arrow[]{>}}}},
    valparticle/.style={postaction={decorate},
    decoration={markings,mark=at position .56 with {\arrow[]{>>}}}}, 
    earlyhole/.style={postaction={decorate},
    decoration={markings,mark=at position .32 with {\arrow[]{<}}}}, 
    hole/.style={postaction={decorate},
    decoration={markings,mark=at position .52 with {\arrow[]{<}}}}, 
    wave/.style={decorate, decoration=snake,thin},
    valhole/.style={postaction={decorate},
    decoration={markings,mark=at position .56 with {\arrow[]{<<}}}}, 
    bparticle/.style 2 args={postaction={decorate},
    decoration={markings,mark=at position .32 with {\arrow[]{>}},mark=at position .72 with {\arrow[]{>}}},
edge node={node [pos=.3,font=\scriptsize] {#1} node [pos=.7,font=\scriptsize] {#2}},
},
    bhole/.style={postaction={decorate},
    decoration={markings,mark=at position .32 with {\arrow[]{<}},mark=at position .72 with {\arrow[]{<}}}}, 
    oneb/.style={minimum width=4mm,path picture={\draw[black,thick]
            (path picture bounding box.south east) -- (path picture bounding box.north west)
            (path picture bounding box.south west) -- (path picture bounding box.north east);
        },
      } 
}

\newcommand{\vertexdistance}{1.2}

\definecolor{indianred}{rgb}{0.86, 0.08, 0.24}
\definecolor{royalblue}{rgb}{0.25, 0.41, 0.88}
\definecolor{darkorange}{rgb}{1.0, 0.55, 0}
\definecolor{mediumseagreen}{rgb}{0.24, 0.70, 0.44}
\definecolor{purple}{rgb}{0.5, 0, 0.5}
\definecolor{cyan3}{rgb}{0, 0.80, 0.80}

\newcommand{\symbolbox}[1][black]{{\color{#1}\scalefont{0.75}$\blacksquare$}}
\newcommand{\symbolcircle}[1][black]{{\color{#1}\scalefont{0.75}\ding{108}}}

\newcommand{\symboldiamondsym}[1][black]{{\color{#1}\scalefont{0.75}\raisebox{-.2ex}{\begin{turn}{45}$\blacksquare$\end{turn}}}}

\newcommand{\bluecircle}{{\scalefont{0.9}\symbolcircle[royalblue]}}

\newcommand{\orangediamond}{{\scalefont{0.9}{\scalefont{0.8}\symboldiamondsym[darkorange]}}}

\newcommand{\redsquare}{{\scalefont{0.9}\symbolbox[indianred]}}

\definecolor{plot1}{rgb}{0.86, 0.08, 0.24}
\definecolor{plot2}{rgb}{0.25, 0.41, 0.88}
\definecolor{plot3}{rgb}{1.0, 0.55, 0}
\definecolor{plot4}{RGB}{61,153,86}

\newcommand{\la}{\langle}
\newcommand{\ra}{\rangle}

\newcommand{\HFB}{\ensuremath{\vert \Phi \ra}}
\newcommand{\Xm}[1]{\ensuremath{#1_\text{max}}}
\newcommand{\Xmin}[1]{\ensuremath{#1_\text{min}}}

\newcommand{\norm}[1]{\left\lVert#1\right\rVert}

\newcommand{\rk}[1]{\ensuremath{r_\text{#1}}}

\usepackage[normalem]{ulem}

\newcommand{\clebsch}[6]{C_{#1 #2 #3 #4}^{#5 #6}}

\newcommand{\sixj}[6]{\begingroup\setlength{\arraycolsep}{0.2em}\begin{Bmatrix} #1 & #2 & #3 \\ #4 & #5 & #6 \end{Bmatrix}_{\text{6j}}\endgroup}

\begin{document}

\title{Pre-processing the nuclear many-body problem}
\subtitle{Importance truncation \emph{versus} tensor factorization techniques}

\author{A.~Tichai\inst{1} \and J.~Ripoche\inst{2} \and T.~Duguet\inst{3,4}}

\institute{
ESNT, CEA-Saclay, DRF, IRFU, D\'epartement de Physique Nucl\'eaire, Universit\'e de Paris Saclay, F-91191 Gif-sur-Yvette, \email{alexander.tichai@cea.fr} \and
CEA, DAM, DIF, F-91297 Arpajon, France, \\ \email{julien.ripoche@cea.fr} \and
IRFU, CEA, Universit\'e Paris-Saclay, 91191 Gif-sur-Yvette, France, \and 
KU Leuven, Instituut voor Kern- en Stralingsfysica, 3001 Leuven, Belgium \\ \email{thomas.duguet@cea.fr}
}

\abstract{
The solution of the nuclear A-body problem encounters severe limitations from the size of many-body operators that are processed to solve the stationary Schr\"odinger equation. These limitations are typically related to both the (iterative) storing of the associated tensors and to the computational time related to their multiple contractions in the calculation of various quantities of interest.  However, not all the degrees of freedom encapsulated into these tensors equally contribute to the description of many-body observables. Identifying systematic and dominating patterns, a relevant objective is to achieve an \emph{a priori} reduction to the most relevant degrees of freedom via a pre-processing of the A-body problem. The present paper is dedicated to the analysis of two different paradigms to do so. The factorization of tensors in terms of lower-rank ones, whose know-how has been recently transferred to the realm of nuclear structure, is compared to a reduction of the tensors' index size based on an importance truncation. While the objective is to eventually utilize these pre-processing tools in the context of non-perturbative many-body methods, benchmark calculations are presently performed within the frame of perturbation theory. More specifically, we employ the recently introduced Bogoliubov many-body perturbation theory that is systematically applicable to open-shell nuclei displaying strong correlations. This extended perturbation theory serves as a jumpstart for non-perturbative Bogoliubov coupled cluster and Gorkov self-consistent Green's function theories as well as to particle-number projected Bogoliubov coupled cluster theory for which the pre-processing will be implemented in the near future. Results obtained in "small" model spaces are equally encouraging for tensor factorization and importance truncation techniques. While the former requires significant numerical developments to be applied in large model spaces, the latter is presently applied in this context and demonstrates great potential to enable high-accuracy calculations at a much reduced computational cost.
}


\maketitle

\section{Introduction} 

Due to the growing amount of data at play in applied and fundamental sciences there exists a strong need for efficient data analysis tools to store and process more efficiently the underlying information. This statement, generally valid independently of the type of data or research area, applies acutely to the nuclear many-body problem that provides a particularly interesting challenge beyond the lightest nuclei.

Over the past decade, tremendous progress have been made to extend the reach of \emph{ab initio} methods to larger mass numbers and better accuracy. In particular the development of many-body methods based on a systematic expansion of the exact solution around a conveniently chosen reference state has allowed an efficient description of medium-mass (semi-)magic nuclei. Examples of such methods are many-body perturbation theory (MBPT)~\cite{Langhammer:2012jx,Hu:2016txm,Tichai:2017rqe,Tichai:2018mll,Arthuis:2018yoo,Hu18arxiv}, self-consistent Green's function (SCGF)~\cite{Dickhoff:2004xx,Soma:2011aj,Soma:2013xha,Carbone:2013eqa,Lapoux:2016exf,Duguet:2016wwr,Raimondi:2017kzi,Raimondi:2018mtv}, coupled-cluster (CC)~\cite{Hagen:2013nca,Signoracci:2014dia,Morris:2017vxi} or the in-medium similarity renormalization group (IM-SRG)~\cite{Hergert:2015awm,Hergert:2016iju,Parzuchowski:2017wcq,Morris:2017vxi} approaches. This category of polynomially-scaling many-body methods is the focus of our attention. 

The guiding idea behind these methods is to account for \textit{dynamical correlations} via a particle-hole (quasi-particle) expansion that is approximated according to a chosen truncation scheme. In the simplest case, low-order MBPT corresponds to keeping the first few terms of an expansion in powers of the residual interaction. In more advanced non-perturbative frameworks, all-order resummations of MBPT contributions are typically accounted for by solving a non-linear set of equations, e.g., the amplitude equations in CC theory. To extend their reach to open-shell nuclei and capture noticeably challenging {\it static} correlations, these single-reference methods employ a symmetry-breaking reference state~\cite{Soma:2011aj,Signoracci:2014dia,Tichai:2018mll} and possibly restore the broken symmetry in a two-step approach~\cite{Duguet:2014jja,Duguet:2015yle,Qiu:2018edx}. 

In all expansion methods, the working equations are eventually expressed as multiple contractions of two different sets of tensors whose indices relate to a given (truncated) basis of the one-body Hilbert space ${\cal H}_1$. The first set of tensors relates to the given of the many-body Hamiltonian
\begin{align}
H_\text{nucl} &= T + V + W +... \label{eq:ham} \\
&\equiv  \frac{1}{(1!)^2} \sum _{pq} t_{pq} c^{\dagger}_{p} c_{q} \nonumber \\ 
&+ \frac{1}{(2!)^2} \sum _{pqrs} \bar{v}_{pqrs}  c^{\dagger}_{p} c^{\dagger}_{q} c_{s} c_{r}   \nonumber \\
&+ \frac{1}{(3!)^2} \sum_{pqrstu} \bar{w}_{pqrstu} c^{\dagger}_{p}c^{\dagger}_{q}c^{\dagger}_{r}c_{u}c_{t}c_{s}  \nonumber \\
&+... \, ,  \nonumber
\end{align}
where $T$ indicates the kinetic energy whereas $V$ and $W$ characterize two- and three-body interactions, respectively. Dots embody possible higher-body operators. The Hamiltonian is thus represented via a set of mode-n (i.e. n-index) tensors, with $n=2, 4$ and $6$ in the above example. The storage cost of the Hamiltonian is thus dominated by the mode-$2k$ tensor defining the highest $k$-body interaction and scales as $N^{2k}$, where $N$ denotes the dimension of the underlying one-body basis used.

The second set of tensors at play is presently denoted as {\it many-body tensors} and is method specific. For example, in the Hartree-Fock- (HF) based CC or MBPT implemented on the basis of a Hamiltonian containing two-body interactions only, the ground-state binding energy can be written as
\begin{align}
E_{0} &= E^{\text{HF}}_{0} + \frac{1}{2} \sum_{ijab} \bar{v}_{ijab} \, t^{a}_{i} \, t^{b}_{i}  + \frac{1}{4} \sum_{ijab} \bar{v}_{ijab} \, t^{ab}_{ij} , \label{CCenergy}
\end{align}
where $t^{a}_{i}$  and $t^{ab}_{ij}$ denote so-called single and double cluster amplitudes. These many-body amplitudes respectively constitute mode-$2$ and mode-$4$ tensors that are contracted with the mode-$4$ tensor originating from the Hamiltonian to compute, e.g., the ground-state energy. Depending on the level at which MBPT or CC are implemented, the calculation of $t^{a}_{i}$ and $t^{ab}_{ij}$ may actually require to carry higher-mode tensors as intermediate objects whose storage scales as $N^{l}$ and whose computation scales as $N^p$. In the non-perturbative CC method, these many-body tensors need to be solved for and stored repeatedly. A similar identification of the relevant many-body tensors can be carried out for SCGF or IMSRG methods.

Eventually, the memory load and the computational cost of a given many-body implementation are respectively driven by the highest mode tensor carried in the calculation and by the complexity of the tensor network associated with the working equations. In non-perturbative approaches, this cost is significantly augmented by the iterative character of the method that requires a repeated computation and storage of possibly high-mode many-body tensors. It is to be noted that the computational load of a given many-body method is significantly more pronounced for nuclei than for, e.g., electronic systems at play in atomic physics or quantum chemistry. This is due (i) to the importance of (at least) three-nucleon interactions, (ii) to significant low-to-high momentum couplings encapsulated into the Hamiltonian tensors and (iii) to the dominance of open-shell nuclei displaying strong correlations. While the first point translates into the necessity to carry (at least) a mode-6 tensor from the outset, the second point renders necessary to use significantly larger one-body bases. The third point makes mandatory to design versatile many-body methods capable of consistently grasping dynamical and static correlations. Eventually, these three features currently forbid the application of polynomially-scaling methods to high accuracy (below 1 per cent error) as well as to nuclei with mass $A>100$ and/or with a doubly open-shell character.

\begin{figure*}[t]
\centering
\scalebox{1.}{
\includegraphics[width=1.3\columnwidth]{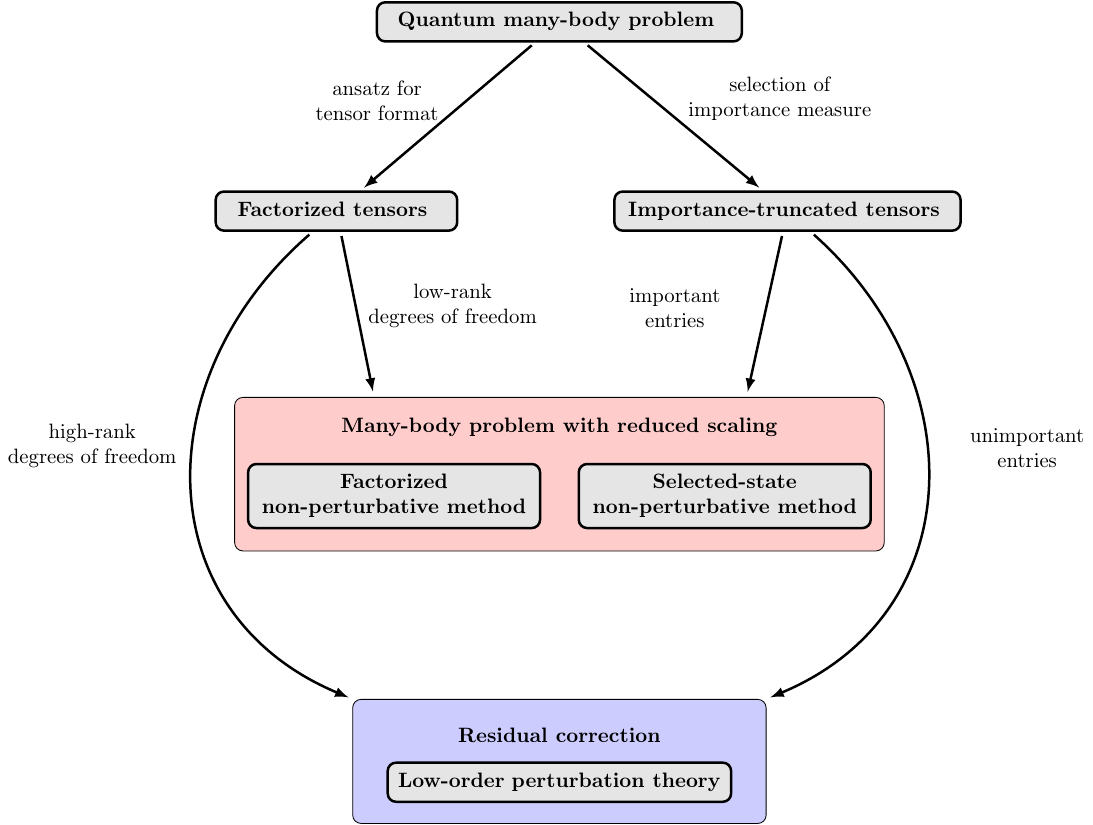}
}
\caption{Pre-processing of the quantum many-body problem on the basis of tensor factorization or importance truncation techniques.}
\label{fig:overview}
\end{figure*}

Generically speaking, overcoming the curse of dimensionality associated with the exponential growth of the $A$-body Hilbert space  is a long-standing problem that has given birth to numerous method- or system-specific strategies. In one way or another, the idea is to exploit an \emph{a priori} knowledge based on physical considerations, educated guess or a simplified evaluation of the importance of selected configurations. This concerns for instance (a) recent advances in tensor network theory (TNT)~\cite{Ver08,Sch11,Or14} where the localization of quantum entanglement over the $A$-body Hilbert space is exploited to describe low-dimensional systems, (b) the recent development of infinite-basis extrapolation techniques in nuclear physics~\cite{FuHa12,Fu14,We15} where an educated guess on the long-distance behavior of the many-body wave function is exploited, or (c) the method at the core of the Monte Carlo shell model~\cite{KoDe97,OtHo01} where the relevance of randomly generated configurations are tested 'on the fly' via an appropriate importance measure.

The present work takes place in this context and wishes to test two systematic strategies appropriate to the many-body methods introduced above. The two types of pre-processing are
\begin{enumerate}
\item \emph{Tensor factorization} (TF) techniques whose goal is to produce a low-rank decomposition of both the Hamiltonian and the many-body tensors at play in the method. The anticipated merit of these techniques is to effectively reduce the value $p$ of the $N^p$ storage and computational costs of the full-fledge implementation.
\item \emph{Importance truncation} (IT) techniques whose goal is to discard entries in the many-body tensors at play in the many-body framework. This is done by estimating the importance of each tensor entry on the basis of a low-order, i.e. computationally cheap, estimate. The anticipated merit of this technique is to effectively reduce the value $N$ of the $N^p$ storage and computational costs of the full-fledge implementation.
\end{enumerate}

Recently, TF techniques have shown to be a promising tool to lower the computational requirements of nuclear many-body calculations~\cite{Tichai:2018eem} dedicated to closed-shell nuclei. Two tensor decomposition formats have been applied to state-of-the-art nuclear Hamiltonians to obtain compressed tensors yielding accurate description of nuclear observables benchmarked within the well-tested second-order Hartree-Fock MBPT (HF-MBPT). The goal of the present paper is to extend this analysis to single-reference methods applicable to open-shell nuclei, i.e. to methods authorizing the spontaneous breaking of symmetries. More specifically, $U(1)$ global-gauge symmetry associated to particle-number conservation is presently authorized to break in order to tackle singly open-shell nuclei. 
This feature is characteristic of Bogoliubov many-body perturbation theory (BMBPT)~\cite{Tichai:2018mll,Arthuis:2018yoo}, particle-number projected BMBPT (PBMBPT)~\cite{Duguet:2015yle}, Bogoliubov CC (BCC) theory~\cite{Signoracci:2014dia,Henderson:2014vka}, particle-number projected BCC (PBCC)~\cite{Duguet:2015yle,Qiu:2018edx} and Bogoliubov-based, i.e. Gorkov, SCGF (GSCGF) theory~\cite{Soma:2011aj,Soma:2013xha}. The present challenge is thus to factorize (i) the Hamiltonian matrix elements expressed in a quasi-particle basis and (ii) the tensor networks associated to the many-body method and truncation level of interest. Regarding point (i), nothing guarantees that the performance of the TF observed in Ref.~\cite{Tichai:2018eem} extends to matrix elements represented in a symmetry-breaking, i.e. Bogoliubov quasi-particle, basis. Regarding point (ii), tests are presently performed using the simplest of methods, i.e. second-order BMBPT, which is thus tensor decomposed and benchmarked against its full-fledged implementation. Extensions to BCC, PBMBPT, PBCC and GSCGF are left to future works.

While TF numerically filters high-rank components of the many-body tensors, another option consists of discarding a large set of their entries on the basis of a robust but inexpensive \emph{a priori} estimate on their importance. Iterative equations are then only solved for the dominant entries while an \emph{a posteriori} correction may be considered for the omitted part of the many-body tensors. As for TF, IT introduces a systematic approximation error that needs to be characterized through benchmark calculations. So far, IT techniques have been mainly used in the context of the no-core shell-model (NCSM) that allows for brute-force solutions of the stationary Schr\"odinger equation up to mass number $A\approx12$. As a remedy to the exponential growth of the $A$-body Hilbert space, the IT-NSCM based on an \emph{a priori} estimate of the relevance of each many-body basis state has allowed to extend the application of diagonalization techniques to $A\approx 25$ in the last decade~\cite{Roth:2007sv,Roth09,Roth:2008qd}. While it is also in connection with configuration interaction (CI) that IT techniques have been mostly developed in quantum chemistry~\cite{buenker74a,buenker75a,illas91a}, it can also be combined with Monte Carlo calculations~\cite{giner13a}. As for non-perturbative expansion techniques, e.g (B)CC, (G)SCGF and IMSRG methods, this has been seldom used. Very recently, an interesting work based on CC theory bearing great resemblance with the presently developed idea, has been realized in quantum chemistry~\cite{deustua17a}. In this scheme, a large part of the triple CC amplitude is omitted on the basis of a prior full configuration interaction quantum Monte Carlo (FCIQMC)~\cite{booth09a,booth10a} or CC Monte Carlo (CCMC)~\cite{thom10a,spencer16a,scott17a} calculation.

The paper is organized as follows. Section~\ref{sec:bmbpt} introduces Bogoliubov coupled cluster theory as the many-body formalism of reference employed throughout the paper to formally illustrate the potential merit of TF and IT techniques. Its perturbative reduction, i.e. Bogoliubov many-body perturbation theory, is further characterized both in view of performing numerical benchmarks and as the inexpensive auxiliary method providing the IT measure. Section~\ref{sec:tensor} provides an overview of TF techniques before displaying the results of their application in small model spaces to quasi-particle matrix elements of the nuclear Hamiltonian. In Sec.~\ref{sec:it}, the IT concept is introduced and applied in small model spaces to be compared to TF techniques. Next, IT results in large model spaces appropriate for medium-mass systems are displayed and analyzed in details. Section~\ref{sec:conclusion} concludes with a set of envisioned extensions for future research.

\section{Many-body formalism}
\label{sec:bmbpt}

The present paper wishes to exemplify the use of TF and IT techniques within the frame of non-perturbative Bogoliubov coupled-cluster theory~\cite{Si15}. The latter thus serves as a baseline for the discussion although actual BCC calculations based on TF and IT are postponed to a future work. In the present paper, the perturbative reduction of BCC, i.e. BMBPT~\cite{Tichai:2018mll,Arthuis:2018yoo}, is employed to perform benchmark calculations.

\subsection{Hamiltonian tensors}

Bogoliubov CC theory and BMBPT expand the exact many-body ground-state around a particle-number breaking Bogoliubov state $\HFB$. The reference state $\HFB$ is a vacuum for a complete set of quasi-particle operators obtained from particle ones through a unitary Bogoliubov transformation
\begin{subequations}
\begin{align}
\beta_k &\equiv \sum_p U^*_{pk} c_p + V^*_{pk} c^\dagger_p\, , \\
\beta_k^\dagger &\equiv \sum_p U_{pk} c^\dagger_p + V_{pk} c_p \, ,
\end{align}
\end{subequations}
such that $\beta_k \HFB =0 $ for all $k$. The columns of the transformation matrices $(U,V)$ are typically obtained as eigenvectors of the HFB equation~\cite{RiSc80} that also delivers the set of quasi-particle energies $\{E_k > 0\}$ as eigenvalues. Since the Bogoliubov reference state is not an eigenstate of the particle-number\footnote{In practice the constraint has to be done for neutron and proton-number operators $N$ and $Z$, respectively by introducing two separate chemical potentials $\lambda_N$ and $\lambda_Z$. In our formalism $A$ stands for either one of them.} operator $A$, the HFB equation is obtained while constraining the expectation value of $A$ in $\HFB$ to match the physical particle number via the use of a Lagrange term, i.e. by minimizing the grand potential
\begin{align}
\Omega \equiv H - \lambda A\, .
\end{align}
The many-body methods of interest are most naturally expressed in terms of normal-ordered contributions to the grand potential $\Omega$ with respect to the Bogoliubov reference state. Starting from the Hamiltonian expressed in the one-body basis (Eq.~\eqref{eq:ham}), one obtains
\begin{align}
\Omega &= \overbrace{\Omega^{00}}^{\displaystyle  \Omega^{[0]}}   \notag \\
&\phantom{=}  + 
 \overbrace{\Omega^{20} + \Omega^{11} +\Omega^{02}}^{\displaystyle  \Omega^{[2]}}  \notag \\
&\phantom{=} +   \overbrace{\Omega^{40} + \Omega^{31} +\Omega^{22} +\Omega^{13} +\Omega^{04}}^{\displaystyle  \Omega^{[4]}}  \notag \\
&\phantom{=}  +  \overbrace{\Omega^{60} + \Omega^{51}  + \Omega^{42} + \Omega^{33} + \Omega^{24} + \Omega^{15} +\Omega^{06}}^{\displaystyle  \Omega^{[6]}}  \, ,
\label{eq:NO}
\end{align}
where $\Omega^{ij}$ denotes the normal-ordered component involving $i$ ($j$) quasi-particle creation (annihilation) operators, e.g., 
\begin{align}
\Omega^{31} &\equiv \frac{1}{3!}\sum_{k_1 k_2 k_3 k_4}  \Omega^{31}_{k_1 k_2 k_3 k_4}
   \beta^{\dagger}_{k_1}\beta^{\dagger}_{k_2}\beta^{\dagger}_{k_3}\beta_{k_4} \, . \label{eq:NOex}
\end{align} 
It is clear from Eqs.~\eqref{eq:NO} and~\eqref{eq:NOex} that the Hamiltonian collects now a set of mode-$(i\!+\!j)$ tensors $\Omega^{ij}_{k_1 \ldots k_{i} k_{i+1} \ldots k_{i+j}}$ that are expressed in a symmetry breaking, i.e. quasi-particle, basis. These tensors display antisymmetry properties, i.e.
\begin{eqnarray}
\Omega^{ij}_{k_1 \ldots k_{i} k_{i+1} \ldots k_{i+j}} &=& (-1)^{\sigma(P)}
\Omega^{ij}_{P(k_1 \ldots k_i | k_{i+1} \ldots k_{i+j})}  \, ,
\end{eqnarray}
where $\sigma(P)$ refers to the signature of the  permutation $P$.  The notation $P(\ldots | \ldots)$ denotes a  separation into the $i$ quasiparticle creation operators and the $j$ quasiparticle annihilation operators such that permutations are only considered among members of the same group. Details on the normal-ordering procedure as well as expressions of the various contributions to $\Omega^{[0]}$, $\Omega^{[2]}$ and $\Omega^{[4]}$ in terms of the original matrix elements of $H$ and of the $(U,V)$ matrices can be found in Ref.~\cite{Si15}. When the Bogoliubov reference state is chosen to solve the HFB equations, one has 
\begin{subequations}
\begin{align}
\Omega^{20}_{k_1 k_2} &=\Omega^{02}_{k_1 k_2}=0 \, , \\
\Omega^{11}_{k_1 k_2} & =E_{k_1}\, \delta_{k_1 k_2} \, .
\end{align} 
\end{subequations}

To circumvent the explicit treatment of three-body operators, i.e., mode-$6$ tensors, state-of-the-art many-body calculations employ the so-called \emph{normal-ordered two-body approximation} (NO2B). In large-scale NCSM calculations, the error induced by the NO2B approximation to the Hamiltonian was estimated to be of the order of $1$-$3\%$~\cite{RoBi12,Geb16} up to the oxygen region. While straightforwardly defined in symmetry-conserving methods, the design of a symmetry-conserving approximation of the Hamiltonian in methods based on a symmetry-breaking reference state is non trivial~\cite{ripoche19a}. Eventually, the approximation leads to omitting $\Omega^{[6]}$ (but not only) such that the dominant contribution of the original three-body interaction is included into the retained terms $\Omega^{[0]}$, $\Omega^{[2]}$ and $\Omega^{[4]}$ via the normal-ordering procedure\footnote{While the terms $\Omega^{[0]}$, $\Omega^{[2]}$ and $\Omega^{[4]}$ are in fact also affected by the approximation~\cite{ripoche19a}, the original notation is kept for simplicity in the remainder of the present paper.}. The present analysis is performed in this context and can later be extended to the use of $\Omega^{[6]}$.

\subsection{Many-body tensors}
\label{Sec.BCC}

Bogoliubov CC theory relies on the use of the ground-state wave-function ansatz~\cite{Si15}
\begin{equation}
\label{e:bccwf}
| \Psi \rangle \equiv e^{\mathcal{T}} | \Phi \rangle \, ,
\end{equation}
where the quasiparticle cluster 
operator $\mathcal{T} \equiv \mathcal{T}_1 +\mathcal{T}_2 + \mathcal{T}_3 + \ldots$ is defined through
\begin{align}
\mathcal{T}_1 &\equiv \frac{1}{2!}\displaystyle\sum_{k_1 k_2}t^{20}_{k_1 k_2} 
\beta^{\dagger}_{k_1}\beta^{\dagger}_{k_2} \, , \nonumber \\
\mathcal{T}_2 &\equiv \frac{1}{4!}\displaystyle\sum_{k_1 k_2 k_3 k_4}t^{40}_{k_1 k_2 k_3 k_4} 
\beta^{\dagger}_{k_1} \beta^{\dagger}_{k_2}\beta^{\dagger}_{k_3} \beta^{\dagger}_{k_4} \, , \\
\mathcal{T}_3 &\equiv \frac{1}{6!}\displaystyle\sum_{k_1 k_2 k_3 k_4 k_5 k_6} 
t^{60}_{k_1 k_2 k_3 k_4 k_5 k_6} \beta^{\dagger}_{k_1} \beta^{\dagger}_{k_2} 
\beta^{\dagger}_{k_3}\beta^{\dagger}_{k_4}\beta^{\dagger}_{k_5}\beta^{\dagger}_{k_6} \, , \nonumber
\end{align}
etc. The BCC amplitudes $t^{m0}_{k_1 \ldots k_m}$, which need to be determined, constitute the fully antisymmetric many-body tensors of present interest.  

In the NO2B approximation, the BCC ground-state energy reads as
\begin{eqnarray}
{\cal E}_{0} &\equiv&  \langle \Phi | \bar{\Omega} | \Phi \rangle \label{e:bccsde} \\
&=& \Omega^{00}\nonumber \\
&& + \frac{1}{2} \sum_{k_1 k_2} \Omega^{02}_{k_1 k_2} \, t^{20}_{k_1 k_2} \nonumber \\
&& + \frac{1}{4!} \sum_{k_1 k_2 k_3 k_4} \Omega^{04}_{k_1 k_2 k_3 k_4} \, t^{40}_{k_1 k_2 k_3 k_4} \nonumber \\
&& + \frac{1}{8} \sum_{k_1 k_2 k_3 k_4} \Omega^{04}_{k_1 k_2 k_3 k_4} \, t^{20}_{k_1 k_2} \, t^{20}_{k_3 k_4} \nonumber \, ,
\end{eqnarray}
where the similarity-transformed grand potential is defined through $\bar{\Omega} \equiv e^{-\mathcal{T}} \Omega e^{\mathcal{T}}$. Similarly to Eq.~\eqref{CCenergy} for standard CC, the BCC correlation energy $\Delta \Omega_0 \equiv {\cal E}_{0} - \Omega^{00}$ is an explicit function of the sole single and double BCC amplitudes. The determination of the BCC amplitudes defining the connected cluster operator $\mathcal{T}$ relies on solving a set of coupled non-linear equations given in a compact form by
\begin{equation}
\label{e:bcceq}
{\cal M}_{k_1 \ldots k_p} \equiv \langle \Phi^{k_1 \ldots k_p} | \bar{\Omega} | \Phi \rangle  =0 \, ,
\end{equation}
where even quasi-particle excitations of the vacuum are defined through
\begin{equation}
\label{e:qpexcit}
| \Phi^{k_1 \ldots k_p}  \rangle  \equiv \beta^{\dagger}_{k_1} \ldots \beta^{\dagger}_{k_p} | \Phi \rangle \, .
\end{equation}

The standard hierarchy of truncation schemes consists in solving the set of equations for $\{\mathcal{T}_1, \ldots,\mathcal{T}_r\}$, i.e. solving Eq.~\eqref{e:bcceq} for $p \leq r$, while setting $\mathcal{T}_q=0$ for $q>r$. For example, BCC with singles and doubles (BCCSD) retains $\mathcal{T}_1$ and $\mathcal{T}_2$ while setting $\mathcal{T}_q=0$ for $q>2$. This corresponds to solving Eq.~\eqref{e:bcceq} in the subspace of Fock space spanned by two and four quasi-particle excitations and denoted as ${\cal F}^{\text{SD}}$ while ignoring $\mathcal{T}_q$ with $q>2$. The explicit tensor network associated with Eq.~\ref{e:bcceq} and used to determine $t^{20}_{k_1 k_2}$ and $t^{40}_{k_1 k_2 k_3 k_4}$ within BCCSD can be found in Ref.~\cite{Si15}. The next approximation level, coined as BCCSDT, consists of further considering triples, i.e. the mode-6 tensor $t^{60}_{k_1 k_2 k_3 k_4 k_5 k_6}$, which impacts the determination of $\mathcal{T}_1$ and $\mathcal{T}_2$, and thus ultimately the energy. Thus, BCCSDT consists of solving Eq.~\eqref{e:bcceq} in the larger subspace of Fock space further spanned by six quasi-particle excitations and denoted as ${\cal F}^{\text{SDT}}$.

\begin{figure}
\begin{center}
\begin{tikzpicture}
\node[draw,omeganode,label=right:$\Omega^{20}_{k_1 k_2}$] (1) at (0,0) {a};
\node (label) at (0,-1) {$t^{20(1)}_{k_1 k_2}$};
\node (2a) at (-0.3,\vertexdistance) {$k_1$};
\node (2b) at (0.3,\vertexdistance) {$k_2$};
\draw (1) edge[particle] node[left] {} (2a);
\draw (1) edge[particle] node[right] {} (2b);
\end{tikzpicture}
\hspace{1.5cm}
\begin{tikzpicture}
\node[draw,omeganode,label=right:$\Omega^{40}_{k_1 k_2 k_3 k_4}$] (1) at (0,0) {};
\node (label) at (0,-1) {$t^{40(1)}_{k_1 k_2 k_3 k_4}$};
\node (2a) at (-0.8,\vertexdistance) {$k_1$};
\node (2b) at (-0.3,\vertexdistance) {$k_2$};
\node (2c) at (0.3,\vertexdistance) {$k_3$}; 
\node (2d) at (0.8,\vertexdistance) {$k_4$}; 
\draw (1) edge[particle] node[left] {} (2a);
\draw (1) edge[particle] node[right] {} (2b);
\draw (1) edge[particle] node[right] {} (2c);
\draw (1) edge[particle] node[right] {} (2d);
\end{tikzpicture}
\end{center}
\caption{Hugenholtz diagrams for the first-order single and double BCC amplitudes.}
\label{fig:wvdiag}
\end{figure}
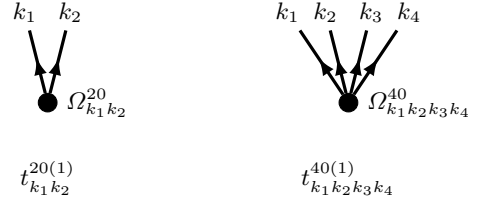

It is possible to move to a perturbative version of BCC, i.e. to employ BMBPT~\cite{Tichai:2018mll,Arthuis:2018yoo}. It consists of bypassing the iterative solving of the amplitudes equations by relying on their perturbative approximations. In the NO2B approximation, cluster amplitudes are given to first order in perturbation by\footnote{Using a HFB reference state leads to $t^{20(1)}_{k_1 k_2}=0$  given that $\Omega^{20}_{k_1 k_2}=0$ in this case. The first non-zero contribution to $t^{20}_{k_1 k_2}$ arises in this case at second order.}
\begin{subequations}
\label{TinBMBPT}
\begin{align}
t_{k_1 k_2}^{20(1)} &= -\frac{\Omega^{20}_{k_1 k_2} }{E_{k_1k_2} } \, , \label{TinBMBPT1} \\
t_{k_1 k_2 k_3 k_4 }^{40(1)} &= -\frac{\Omega^{40}_{k_1 k_2 k_3 k_4} }{E_{k_1k_2k_3k_4}} \, ,  \label{TinBMBPT2} \\
t_{k_1 k_2 k_3 k_4 k_5 k_6}^{60(1)} &= 0 \, , \label{TinBMBPT3}
\end{align}
\end{subequations}
where 
\begin{align}
E_{k_1 k_2 k_3 k_4 \ldots} &\equiv E_{k_1} + E_{k_2} + E_{k_3} + E_{k_4}+\ldots \, .
\end{align}
The two non-zero contributions in Eq.~\eqref{TinBMBPT} correspond to the Hugenholtz diagrams displayed in Fig.~\ref{fig:wvdiag}~\cite{Si15}. Inserting these expressions in Eq.~\eqref{e:bccsde} while omitting the last term provides the second-order BMBPT correlation energy $\Delta \Omega_0^{(2)}$.

\subsection{$J$-coupled scheme}
\label{Sec.Jcoupled}

Present calculations are performed while imposing spherical symmetry.  One starts from the $N$ lowest one-body eigenstates of the spherical harmonic oscillator (HO) whose associated quantum numbers are
\begin{align}
k \equiv (n_k, l_k, j_k, m_k, t_k) \, ,
\end{align}
where $n_k$ denotes the radial HO quantum number, $l_k$ the orbital angular-momentum quantum number, $j_k$ the total angular-momentum quantum number, $m_k$ the angular momentum projection and $t_k$ the isospin projection. Proceeding to the $J$-coupling of cross-coupled~\cite{Tichai19unp} grand potential and BCC amplitude matrix elements, one obtains the $J$-coupled form of the mode-4 tensors, e.g.
  \begin{align}
{^J \tilde{\Omega}}^{ij}_{\tilde k_1 \tilde k_2 \tilde k_3 \tilde k_4} \equiv \la \tilde k_1 \tilde k_2 (J) | \tilde \Omega^{ij} | \tilde k_3 \tilde  k_4 (J) \ra\, ,
\label{eq:inttensJ}
\end{align}
where each index denotes now the reduced set of quantum numbers
\begin{align}
\tilde k \equiv (n_k, l_k, j_k, t_k) \, ,
\end{align}
differing from $k$ by the removal of the magnetic quantum number $m_k$. Correspondingly, one initial mode-$n$ tensor\footnote{For mode-6 tensors the $J$-coupling is such that the first and second (fourth and fifth) individual angular momenta are coupled to a given $J_{12}$ ($J_{45}$), which is further coupled with the third (sixth) individual angular momentum to a good overall diagonal $J$. See App.~\ref{mode6coupling} for details.} generates a set of $J$-coupled tensors associated with all possible values of the recoupled $(n/2)$-body angular momentum $J$.

Starting from the original tensor indices $\{k_i\}$ whose range $N$ is governed by the truncation of the HO one-body basis characterized by the maximum value of the quantum number $e_\text{max} = 2n_k +l_k$, the range of the indices $\{\tilde{k}_i\}$ of the $J$-coupled tensors is reduced to $\tilde{N}$.

Correspondingly, the tensor networks at play in the many-body method of interest are processed via angular momentum coupling techniques, e.g. the second term in the BCC correlation energy (Eq.~\ref{e:bccsde}) is re-expressed as
\begin{equation}
\label{e:bccsdeJ}
\frac{1}{4!}  \sum_{J} \hat{J}^2 \!\!\! \sum_{\tilde k_1 \tilde k_2 \tilde k_3 \tilde k_4} \, ^J{\tilde{\Omega}^{04}_{\tilde k_1 \tilde k_2 \tilde k_3 \tilde k_4}} \, ^J{\tilde{t}^{40}_{\tilde k_1 \tilde k_2 \tilde k_3 \tilde k_4}}  \, ,
\end{equation}
where $\hat x \equiv \sqrt{2x+1}$~\cite{VaMo88}. Similarly, the second-order BMBPT correlation energy based on a HFB reference state reduces to
\begin{equation}
\label{e:BMBPTdeJ}
\Delta\Omega^{(2)}_0 =  \frac{1}{4!} \sum_{J} \hat{J}^2 \!\!\! \sum_{\tilde k_1 \tilde k_2 \tilde k_3 \tilde k_4}  \, ^J{\tilde{\Omega}^{04}_{\tilde k_1 \tilde k_2 \tilde k_3 \tilde k_4}} \, ^J{\tilde{t}^{40(1)}_{\tilde k_1 \tilde k_2 \tilde k_3 \tilde k_4}}  \, ,
\end{equation}
where the $J$-coupled form of the first-order approximation to the double amplitudes (Eq.~\ref{TinBMBPT2}) reads as\footnote{Working with a spherically-restricted Bogoliubov state, HFB quasi-particle energies are $m_k$-independent, i.e., $E_{\tilde k} = E_k$.}
\begin{equation}
^J{\tilde{t}^{40(1)}_{\tilde k_1 \tilde k_2 \tilde k_3 \tilde k_4}} =  -\frac{^J{\tilde{\Omega}^{40}_{\tilde k_1 \tilde k_2 \tilde k_3 \tilde k_4}}}{E_{\tilde k_1 \tilde k_2 \tilde k_3 \tilde  k_4}}  \, . \label{1storderT2}
\end{equation}

\subsection{Dimensionalities}
\label{Sec.dimensionalities}

\begin{table*}[t!]
\def\arraystretch{1.6}
\centering
\begin{tabular}{c | c | c | cc | cc | cc | c c}
\hline \hline 
& \multicolumn{2}{c |}{Index size} &   \multicolumn{4}{  c  |}{ ${^J \tilde{T}}^{40}_{\tilde k_1 \tilde k_2 \tilde k_3 \tilde k_4}$  }  & \multicolumn{4}{ c }{${^{J_{12} J_{45} J} \tilde{T}}^{60}_{\tilde k_1 \tilde k_2 \tilde k_3 \tilde k_4 \tilde k_5 \tilde k_6}$  } \\ 
\hline 
$e_\text{max}$ & $N$ & $\tilde N$ & \multicolumn{2}{c}{Naive}  & \multicolumn{2}{c|}{Optimal} &  \multicolumn{2}{c}{Naive}   & \multicolumn{2}{c}{Optimal}   \\
\hline \hline
2 & 40 & 12 & $1.2\cdot10^5$ & $1.0$ Mb & $6.5\cdot10^2$ & $5$ kb & $4.3\cdot10^8$ & $3.1$ Gb & $2.2\cdot10^5$ & $0.1$ Mb \\
4 & 140 & 30 & $8.1\cdot10^6$ & $61.8$ Mb & $1.9\cdot10^4$ & $0.1$ Mb & $7.0\cdot10^{11}$ & $5.0$ Tb &     $7.3\cdot10^6$ & $54.4$ Mb \\
6 & 336 & 56 & $1.4\cdot10^8$ & $1.0$ Gb & $2.4\cdot10^5$ & $1.8$ Mb  & $1.6\cdot10^{14}$ & $1.1\cdot10^3$ Tb & $5.0\cdot10^8$ & $3.8$ Gb \\
8 & 660 & 90 & $1.2\cdot10^9$ & $8.8$ Gb & $1.7\cdot10^6$ & $13.0$ Mb & $1.5\cdot10^{16}$ & $1.1\cdot10^6$ Tb &      $1.4\cdot10^{10}$ &  $103.1$ Gb \\
10 & 1140 & 132 & $6.7\cdot10^9$ & $49.8$ Gb & $8.7\cdot10^6$ & $66.0$Mb & $6.9\cdot10^{17}$ & $5.0\cdot10^{7}$ Tb & $2.1\cdot10^{11}$ & $1.5$ Tb \\
12 & 1820 & 182 & $2.9\cdot10^{10}$ & $212.5$ Gb & $3.4\cdot10^7$ &   $261.5$ Mb & $9.2\cdot10^{18}$ & $6.7\cdot10^{8} $ Tb & $2.1\cdot10^{12}$ & $15.2$ Tb \\
\hline \hline
\end{tabular}
\caption{Number of entries and associated memory of $J$-coupled mode-4 and mode-6 tensors expressed in the $U(1)$-breaking quasi-particle basis as a function of the truncation parameter $e_\text{max}$ of the initial one-body spherical HO basis. Two storage schemes are employed (see text). Storage estimates assume double precision for all tensor entries. The quoted numbers correspond to the complete tensor, i.e. they sum contributions from all possible $J$ ($J_{12},J_{45},J$) blocks.}
\label{tabledimensions}
\end{table*}

In order to anticipate the benefit of pre-processing the solving of the $A$-body Schr\"odinger equation, let us now briefly discuss typical dimensionalities and memory requirements. Table~\ref{tabledimensions} provides the numbers of entries and associated memory of $J$-coupled mode-4 and mode-6 tensors expressed in the $U(1)$-breaking quasi-particle basis as a function of the truncation parameter $e_\text{max}$ of the initial spherical HO basis. To this truncation parameter correspond the basis dimension $N$ of the spherical HO basis and the reduced range $\tilde{N}$ of the indices actually labelling the $J$-coupled tensors. Numbers quoted in Tab.~\ref{tabledimensions} correspond to the full mode-4 (mode-6) tensor, i.e. they sum the contributions from all $J$ ($J_{12},J_{45},J$) blocks generated through the angular momentum recoupling. For further orientation, Tab.~\ref{Jblocks} provides the number of those summed blocks as a function of $e_\text{max}$.  The mode-4 (mode-6) tensor under consideration in Tab.~\ref{tabledimensions} is representative of $\Omega^{40}$, $\Omega^{04}$ or ${\cal T}_2$ ($\Omega^{60}$, $\Omega^{06}$ or ${\cal T}_3$) but numbers would be similar for the other contributions to $\Omega^{[4]}$ ($\Omega^{[6]}$), e.g. $\Omega^{22}$ ($\Omega^{33}$).

\begin{table}[t!]
\def\arraystretch{1.6}
\centering
\begin{tabular}{c  | c | c | c | c}
\hline \hline 
&   \multicolumn{2}{  c  |}{ ${^J \tilde{T}}^{40}_{\tilde k_1 \tilde k_2 \tilde k_3 \tilde k_4}$  }  & \multicolumn{2}{ c }{ ${^{J_{12} J_{45} J} \tilde{T}}^{60}_{\tilde k_1 \tilde k_2 \tilde k_3 \tilde k_4 \tilde k_5 \tilde k_6}$  } \\ 
\hline
$e_\text{max}$  & Naive  & Optimal &  Naive & Optimal   \\
\hline \hline
2  & 6 & 6 & 288 & 75 \\
4  & 10 & 10 & 1400 & 289 \\
6  & 14 & 14 & 3920 & 727 \\
8  & 18 & 18 & 8424 & 1469 \\
10  & 22 & 22 & 15488 & 2595 \\
12  & 26 & 26 & 25688 & 4185 \\

\hline \hline
\end{tabular}
\caption{Number of different angular-momentum blocks of $J$-coupled mode-4 and mode-6 tensors in the two storage formats.}
\label{Jblocks}
\end{table}

Two storage schemes are considered. First, $J$-coupled tensors are naively stored with all $\{\tilde{k}_i\}$ indices running independently over the $\tilde{N}$ possible sets of reduced quantum numbers. Second, all symmetries of the $J$-coupled tensors, i.e. antisymmetry under the exchange of pair of indices, parity, isospin\footnote{The coupling to two-body isospin cannot be exploited for quasi-particle matrix elements. The only constraint left from isospin symmetry is that an even number of indices must carry a neutron or a proton label.} and triangular inequalities associated with the (successive) recoupling of pairs of angular momenta, are exploited to avoid storing a large set of null entries. This defines the optimal storage scheme.

While proof-of-principle calculations using $e_\text{max} = 4$ are performed in Secs.~\ref{sec:tensor} and~\ref{ITsmallscale}, converged calculations of mid-mass nuclei with $A\sim 40-80$ whose results are discussed in Sec.~\ref{Sec.largescale} typically require $e_\text{max} = 12$ when using a Hamiltonian softened via a Similarity Renormalization Group (SRG) transformation as described below. Even higher values of $e_\text{max}$ are necessary to employ Hamiltonians that are not processed via SRG and/or to compute heavier nuclei. While today's high performance computers may allow up to 3TB of RAM for a designated memory node and the use of MPI parallelisation may allow to go beyond that, as a rule of thumb, one may consider useful to work with tensors requiring less than 200GB of storage in production calculations\footnote{Non-perturbative methods typically require to store several copies of the same tensor produced through successive iterations.}. Table~\ref{tabledimensions} demonstrates that, while an optimal storage scheme makes the handling of $J$-coupled mode-4 tensors not problematic for $e_\text{max} = 12$, the use of $J$-coupled mode-6 tensors, e.g. the handling of triples in BCC theory, is already a tremendous task. Everything becomes all the more challenging when working in $m$-scheme to, e.g., further authorize rotational symmetry to address doubly open-shell nuclei. The above numbers justify to investigate systematic methods to bypass the full handling of large tensors for a (hopefully) negligeable loss of accuracy.

\subsection{Hamiltonian}
\label{Hamiltonian}

The nuclear Hamiltonian (Eq.~\eqref{eq:ham}) employed in this work has been derived within the frame of chiral effective field theory~\cite{We90,We91,Ep09}. It combines a chiral two-nucleon interaction at next-to-next-to-next-to leading order with a cutoff of $\Lambda_{2N}=500 \,\text{MeV}$~\cite{EnMa03} with a three-nucleon interaction at next-to-next-to leading order with a local regulator based on a cutoff of $\Lambda_{3N}=400 \,\text{MeV}$~\cite{Na07,Roth:2011vt}.

The Hamiltonian is further softened using a SRG transformation with a flow parameter $\alpha=0.08\,\text{fm}^4$~\cite{BoFu07,HeRo07,RoRe08,RoLa11,JuMa13}. This transformation induces many-nucleon forces that are included consistently up to the three-nucleon level, i.e., chiral and induced many-body forces beyond that level are neglected. SRG-evolved Hamiltonians have already been used in a number of medium-mass calculations and have been shown to be soft enough to be used meaningfully in MBPT~\cite{Tichai:2016joa,Tichai:2017rqe,Hu16} and BMBPT~\cite{Tichai:2018mll} calculations.

\section{Tensor factorization} 
\label{sec:tensor}

In Ref.~\cite{Tichai:2018eem}, TF techniques have been used for the first time in the context of \emph{ab initio} nuclear structure calculations. In such an approach, tensors are decomposed and approximated on the basis of a particular tensor format.

\subsection{Basic considerations}

A mode-$n$ tensor $T$ is a multivariate data array $T_{i_1...i_n}$ with index ranges $\{I_1,...,I_n\}$ and can be seen as a higher-mode analogue of vectors and matrices. In order to quantify the distance between tensors, the \emph{Frobenius norm} of a tensor $T$ is introduced as
\begin{align}
\norm{ T } \equiv  \sqrt{\sum_{i_1...i_n} T_{i_1 ... i_n} T^\ast_{i_1 ... i_n}}\, .
\end{align}
Given an approximation $\hat T$ of $T$, the \emph{relative error} is thus defined as
\begin{align}
\Delta T \equiv \frac{ \norm{ T - \hat T }}{\norm{ T}} \, .
\end{align}

\subsection{Tensor hypercontraction}

The key idea behind a given approximation is the decomposition of a mode-$n$ tensor into a sum of products of mode-$k$ tensors with $k<n$.  The particular topology of the decomposition is referred to as a \emph{tensor format}. Over the past years various tensor formats have been successfully applied in atomic and solid-state physics as well as in quantum chemistry, e.g., canonical polyadic decomposition (CPD), resolution of identity (RI) or tensor hypercontraction (THC).

Tensor hypercontraction~\cite{Ho12a,Ho12b,Schu17} consists in expanding a mode-4 tensor under the particular form
\begin{align}
T_{i_1 i_2 i_3 i_4}  = \sum_{\alpha \beta} X^1_{i_1 \alpha} X^2_{i_2 \alpha} W_{\alpha \beta} X^3_{i_3 \beta} X^4_{i_4 \beta}\, .
\label{eq:thc}
\end{align}
Every tensor format comes with a \emph{tensor rank} that is defined as the size of the auxiliary index used when truncating the decomposition. In the present case, the THC rank $\rk{THC}$ is taken to be the same for both summations over $\alpha$ and $\beta$ in Eq.~\eqref{eq:thc}. Increasing the tensor rank of a given decomposition lowers the error such that a systematically improvable approximation is obtained. The THC decomposition involves five factor matrices: four factors $\{X^i\}$ whose size scales linearly with $\rk{THC}$ and the core tensor $W$ that scales quadratically with $\rk{THC}$.

The THC format is very flexible and encompasses the CPD or the purely separable ansatz as particular cases, i.e. setting $W_{\alpha \beta}=\delta_{\alpha\beta}$ the CPD format is recovered as
\begin{align}
T_{i_1 i_2 i_3 i_4}  = \sum_{\alpha} X^1_{i_1 \alpha} X^2_{i_2 \alpha} X^3_{i_3 \alpha} X^4_{i_4 \alpha}\, ,
\label{eq:cpd}
\end{align}
whereas further assuming $\rk{THC}=1$ provides the purely separable approximation
\begin{align}
T_{i_1 i_2 i_3 i_4}  \approx  X^1_{i_1} X^2_{i_2} X^3_{i_3} X^4_{i_4}\, .
\label{eq:sep}
\end{align}

Of importance is the \emph{data compression ratio} 
\begin{align}
R_c \equiv \frac{\text{storage required for $T$}}{\text{storage required for $\hat T$}} \, ,
\end{align}
which relates the initial amount of data in the optimal storage scheme to the compressed amount of data after the decomposition process and the truncation have been achieved. Whenever $R_c>1$, the compressed tensor $\hat T$ requires less storage than the original one.

\subsection{Factorized grand potential}

The algorithmic procedure to determine the THC factor matrices was laid out in Ref.~\cite{Tichai:2018eem}. A numerical analysis of each of the individual steps was provided through the application of THC to state-of-the-art 2N+3N nuclear Hamiltonians represented in various symmetry-conserving bases. The same procedure is presently employed to decompose each component\footnote{It is only necessary to investigate $^J{\Omega^{22}}$, $^J{\Omega^{31}}$ and $^J{\Omega^{40}}$ given that matrix elements of $^J{\Omega^{13}}$ and $^J{\Omega^{04}}$ are trivially related to those of $^J{\Omega^{31}}$ and $^J{\Omega^{40}}$~\cite{Si15}.} of $\Omega^{[4]}$ in $J$-coupled scheme.

\begin{figure}[t!]
\includegraphics{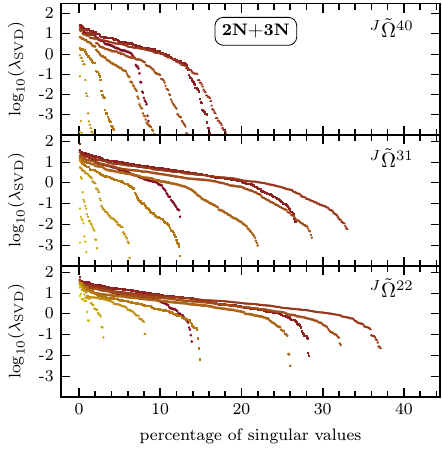}
\caption{(Color online) Singular values for the three normal-ordered components of $\Omega^{[4]}$ in $^{18}$O. Calculations are performed using an $\Xm{e}=4$ HO single-particle basis. In each panel, increasing values of $J$ correspond to decreasingly darker curves.}
\label{fig:svd}
\end{figure}

The first step of the THC decomposition involves a (truncated) singular value decomposition (SVD). Grouping the first and second pair of indices\footnote{Given that quasi-particle matrix elements do not necessarily have the same number of creator and annihilation operators, there is no \emph{a priori} natural coupling between the first two and the last two labels. However, the step leading to cross-coupled matrix elements~\cite{Tichai19unp} does provide a natural coupling, which is indeed used here.}, the mode-4 tensors are rewritten as
\begin{align}
^J{\tilde{\Omega}^{ij}_{\tilde k_1 \tilde k_2 \tilde k_3 \tilde k_4}} \equiv {^J} \tilde{\Omega}^{ij}_{\tilde{K}\tilde{K}'} \, ,
\end{align}
where $\tilde{K}\equiv(\tilde k_1,\tilde k_2)$ and $\tilde{K}'\equiv(\tilde k_3,\tilde k_4)$ are two-body indices of range $\tilde{N}^2$.  The SVD is applied to  $^J \tilde{\Omega}^{40}_{\tilde{K}\tilde{K}'}, ^J \tilde{\Omega}^{31}_{\tilde{K}\tilde{K}'}$ and $^J \tilde{\Omega}^{22}_{\tilde{K}\tilde{K}'}$ and the absolute sizes of the (ordered) sequence of singular values are displayed in Fig.~\ref{fig:svd}. Calculations are performed in $^{18}$O using $e_\text{max}=4$. As visible in Tab.~\ref{tabledimensions}, the corresponding one-body index range is $N=140$ for the original $m$-scheme matrix elements and $\tilde{N}=30$ for the $J$-coupled matrix elements under present investigation.

\begin{figure}
\centering
\includegraphics[width=0.9\columnwidth]{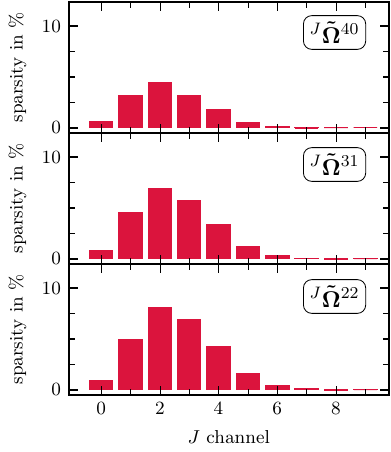}
\caption{(Color online) Histogram of the percentage of non-zero entries of the three components of $\Omega^{[4]}$ as a function of $J$. Calculations are performed in $^{18}$O starting from an $e_\text{max}=4$-truncated one-body HO basis.}
\label{fig:sparsityJ}
\end{figure}

Singular values behave differently for the three grand potential components. The most efficient truncation can be achieved for $^J \tilde{\Omega}^{40}$ for which only 20\% of singular values need to be retained even for intermediate $J$ values. In the most optimal $J$ blocks, up to 95\% of the singular values can be safely discarded. The situation is slightly worse for $^J \tilde{\Omega}^{31}$ and $^J \tilde{\Omega}^{22}$ that require keeping up to 35\% of the singular values, although large $J$ values, again, authorize more severe truncations. As Fig.~\ref{fig:sparsityJ} testifies, these results directly correlate with the sparsity of each tensor ${^J} \tilde{\Omega}^{ij}_{\tilde{K}\tilde{K}'}$, i.e. the smaller the number of initial non-zero entries, the faster the decrease of the singular values. 

While not shown here, similar conclusions have been reached for other semi-magic nuclei. Furthermore, results are qualitatively and quantitatively similar to those obtained for matrix elements of the Hamiltonian represented in the symmetry-conserving HF basis in closed-shell nuclei~\cite{Tichai:2018eem}. As a matter of fact, singular values presently extracted for $^J \tilde{\Omega}^{40}$ decrease even faster. Inspecting the expression of $\Omega^{ij}_{k_1 k_2 k_3 k_4}$ in terms of the Bogoliubov matrices $(U,V)$~\cite{Si15}, such a positive outcome could not be easily anticipated. Indeed, the grouping of quasiparticle indices in  ${^J} \tilde{\Omega}^{ij}_{\tilde{K}\tilde{K}'}$ relates to a mixture of so-called natural and un-natural grouping of single-particle indices in the original matrix elements~\cite{Tichai:2018eem}. This feature directly reflects the presence of pairing correlations. Knowing that the un-natural grouping of single-particle indices was shown to lead to very slowly decreasing singular values and inefficient truncations~\cite{Tichai:2018eem}, the results displayed in Fig.~\ref{fig:svd} constitute a non-trivial result that need to be further investigated in heavier nuclei that are more strongly paired than the oxygen isotopes. 

\begin{figure}[t!]
\includegraphics{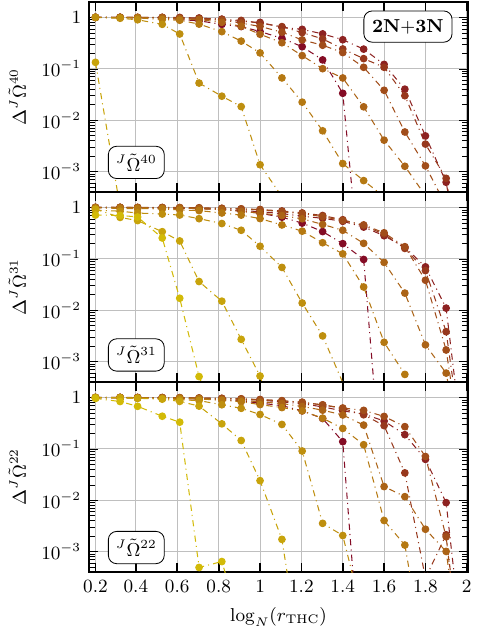}
\caption{Relative error of the THC decomposition  for the three normal-ordered components of $\Omega^{[4]}$ in $^{18}$O. All calculations are performed using an $\Xm{e}=4$ model space. In each panel, increasing $J$ values correspond to decreasingly darker curves.}
\label{fig:THCrank}
\end{figure}

\begin{figure*}[t!]
\centering
\includegraphics{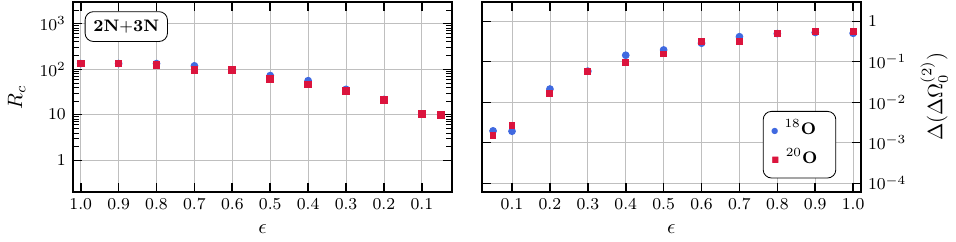}
\caption{Compression ratio and second-order BMBPT corrections for $^{18,20}$O as a function of decomposition error $\epsilon$ . All calculations are performed using an $\Xm{e}=4$ model space.}
\label{fig:bmbpt2}
\end{figure*}

Figure~\ref{fig:THCrank} shows the THC error as a function of the THC rank $r_\text{THC}\equiv\tilde{N}^\alpha$. The THC rank is expressed as a certain power $\alpha$ of the range of the tensor indices to emphasize the relation to the basis dimension. The behavior of the decomposition error as a function of the two-body angular-momentum $J$ is the same for the three grand potential components, i.e., for high values of $J$ the approximation error drops more rapidly than for lower values of $J$. For all $J$ values, a THC approximation error $\Delta ^J{\tilde{\Omega}^{ij}}=10^{-1}$ can be obtained for $\rk{THC}=\tilde{N}^{1.6}$ for $^J{\tilde{\Omega}^{40}}$ and for $\rk{THC}=\tilde{N}^{1.8}$ for $^J{\tilde{\Omega}^{31,22}}$, thus, allowing for a robust extension of the THC ansatz to quasi-particle matrix elements.

To illustrate the memory gain obtained via the application of THC, the data compression factor $R_C$ is displayed in the left panel of Fig.~\ref{fig:bmbpt2} for $^{18,20}$O. The approximation on the overall $\Omega^{[4]}$ tensor is set by using an error threshold $\epsilon$ common to all $J$ channels. This corresponds to using $J$- and component-dependent THC ranks obtained by setting $\Delta ^J{\tilde{\Omega}^{ij}} = \epsilon$ for all $(i,j)$ and $J$. While $R_C=1$ in the limit where no approximation is made on the grand potential tensors, the compression factor increases with decreasing accuracy, i.e. increasing $\epsilon$.  The trend is monotonous and identical for both nuclei under investigation. Eventually, setting $\epsilon=10^{-1}$ authorizes to compress the data by about a factor of 10. While this result is achieved in a small model space, much larger compression factor (for the same error threshold) are envisioned for larger model spaces appropriate to realistic \emph{ab initio} calculations.

\subsection{Factorized many-body tensor and energy}

Eventually, the goal is to decompose both Hamiltonian and many-body tensors at play, eventually leading to a tensor-factorized form of the many-body formalism~\cite{Schu17}. This is envisioned to be truly beneficial in non-perturbative methods where the iterative equations  can be solved for the low-rank factors in a specific tensor format rather than for the original many-body tensors. While this is our goal to do so for, e.g., BCC in the future, second-order BMBPT introduced in Secs.~\ref{Sec.BCC} and~\ref{Sec.Jcoupled} is presently employed to gauge the propagated error to the ground-state energy. 

The decomposition of the tensor network associated with $\Delta\Omega^{(2)}_0$ can be achieved by inserting the decomposition of $^J{\tilde{\Omega}^{04}_{\tilde k_1 \tilde k_2 \tilde k_3 \tilde k_4}}$ and $^J{\tilde{t}^{40(1)}_{\tilde k_1 \tilde k_2 \tilde k_3 \tilde k_4}}$ into Eq.~\eqref{e:BMBPTdeJ}. The THC decomposition of the grand potential tensor is already at hands under the form
\begin{align}
^J{\tilde{\Omega}^{04}_{\tilde k_1 \tilde k_2 \tilde k_3 \tilde k_4}}  = \sum_{\alpha \beta} {^J}{X^1_{\tilde k_1 \alpha}} \, ^J{X^2_{\tilde k_2 \alpha}} \, ^J{W_{\alpha \beta}} \, ^J{X^3_{\tilde k_3 \beta}} \, ^J{X^4_{\tilde k_4 \beta}}\, .
\label{eq:thcOmega40}
\end{align}
A decomposition\footnote{An alternative would be to perform the numerical THC decomposition of $^J{\tilde{t}^{04(1)}_{\tilde k_1 \tilde k_2 \tilde k_3 \tilde k_4}}$ itself rather than combining the decompositions of the tensors that compose it as is presently done. In an actual non-perturbative BCC calculation, an ansatz for the factorization of the cluster amplitudes would be made a priori.} of $^J{\tilde{t}^{40(1)}_{\tilde k_1 \tilde k_2 \tilde k_3 \tilde k_4}}$ (Eq.~\ref{1storderT2}) is achieved by combining Eq.~\eqref{eq:thcOmega40} with the inverse Laplace transform of the energy factor in Eq.~\ref{1storderT2}
\begin{align}
\frac{1}{E_{\tilde k_1\tilde k_2\tilde k_3\tilde k_4} } &= \int_0^\infty e^{-tE_{\tilde k_1\tilde k_2\tilde k_3\tilde k_4} } dt \nonumber \\
& \approx \sum_{s=-M}^{+M} \omega_s e^{-t_s(E_{\tilde k_1} + E_{\tilde k_2} + E_{\tilde k_3} + E_{\tilde k_4} )} \nonumber \\
&\equiv \sum_{s=-M}^{+M}   \pi_{\tilde k_1s} \, \pi_{\tilde k_2s} \, \omega_s \, \pi_{\tilde k_3 s} \, \pi_{\tilde k_4 s} \, , 
\label{eq:dencpd}
\end{align}
where a numerical quadrature has been used. Equation~\eqref{eq:dencpd} corresponds to a CPD whose rank is $r_D\equiv 2M+1$. Numerical values of $M$, $\omega_s$ and $t_s$ are tabulated in the literature~\cite{BrHa05} such that high precision can be reached with very few grid points. In particular, the size of the integration mesh was shown to be independent of the system. The pairing gap associated with the spontaneous breaking of particle-number symmetry makes all energy denominators non-vanishing and, therefore, the quadrature well-defined\footnote{If $U(1)$ symmetry does not spontaneously break, HFB quasi-particle energies reduce to HF single-particle energies with a significant shell gap near the Fermi energy, i.e. a significant HOMO-LUMO gap in the quantum-chemistry language. Consequently, the quadrature is also well defined in this particular limit.}. Combining the above elements, the factorized form of the second-order BMBPT energy correction is obtained as
\begin{align}
\Delta\Omega^{(2)}_0 &= 
-\frac{1}{4!} \sum_{J \alpha \beta  \gamma \delta s} \hspace{-4pt}\hat J^2 
{^{J\hspace{-2pt}}  W}_{\alpha \beta} 
{^{J\hspace{-2pt}}  W}_{\gamma \delta}
{^{J\hspace{-2pt}} A}^{\alpha \gamma}_s
{^{J\hspace{-2pt}} B}^{\alpha \gamma}_s
{^{J\hspace{-2pt}} C}^{\beta \delta}_s
{^{J\hspace{-2pt}} D}^{\beta \delta}_s \, , \nonumber
\end{align}
where the intermediates
\begin{subequations}
\allowdisplaybreaks
\begin{align}
{^{J\hspace{-2pt}} A}^{\alpha \gamma}_s &\equiv \sum_{\tilde k} {^J X}^1_{\tilde k \alpha} {^J X}^1_{\tilde k \gamma} \pi_{\tilde k s} \,, \\
{^{J\hspace{-2pt}} B}^{\alpha \gamma}_s &\equiv \sum_{\tilde k} {^J X}^2_{\tilde k \alpha} {^J X}^2_{\tilde k \gamma} \pi_{\tilde k s} \, ,\\
{^{J\hspace{-2pt}} C}^{\beta \delta}_s &\equiv \sum_{\tilde k} {^J X}^3_{\tilde k \beta} {^J X}^3_{\tilde k \delta} \pi_{\tilde ks} \, , \\
{^{J\hspace{-2pt}} D}^{\beta \delta}_s &\equiv \sum_{\tilde k} {^J X}^4_{\tilde k \beta} {^J X}^4_{\tilde k \delta} \pi_{\tilde ks} \, ,
\end{align}
\end{subequations}
have been introduced. The evaluation cost of these intermediates is  $\mathcal{O} (r^2_\text{THC}   N  r_D)$. Defining new intermediates via
\begin{subequations}
\begin{align}
{^{J\hspace{-2pt}} M}^{\alpha \delta}_s   &\equiv \sum_{\gamma}{^{J\hspace{-2pt}} A}^{\alpha \gamma}_s  {^{J\hspace{-2pt}} B}^{\alpha \gamma}_s  {^{J\hspace{-2pt}}  W}_{\gamma \delta} \, , \\
{^{J\hspace{-2pt}} N}^{\alpha \delta}_s  &\equiv \sum_{\beta} {^{J\hspace{-2pt}} C}^{\beta \delta}_s  {^{J\hspace{-2pt}} D}^{\beta \delta}_s {^{J\hspace{-2pt}}  W}_{\alpha \beta}  \, ,
\end{align}
\label{eq:r3int}
\end{subequations}
one is eventually left with
\begin{align}
\Delta \Omega^{(2)}_0 = -\frac{1}{4!} \sum_{J} \hspace{-2pt} \hat J^2 \, \sum_s \sum_{\alpha \delta} {^{J\hspace{-2pt}} M}^{\alpha \delta}_s {^{J\hspace{-2pt}} N}^{\alpha \delta}_s  \, . \label{finalMBPT2}
\end{align}

The computational scaling of the factorized $\Delta\Omega^{(2)}_0$ depends on the THC ranks $r_\text{THC}$ in each $J$ block, which themselves depend on the chosen approximation error $\epsilon$ on the grand potential. As discussed in Ref.~\cite{Tichai:2018eem}, the scaling is typically worse than the naive $\mathcal{O}(N^4)$ scaling of the original second-order correction. However, the aim of the present work is not to derive a low-scaling approximation of the already low-cost BMBPT(2) but rather to benchmark the propagation of the TF error to nuclear observables. One has to move to many-body methods that are more expensive to begin with to generate a reduction of the numerical scaling.

To measure the impact of TF, the relative error on the second-order energy correction is introduced
\begin{align}
\Delta (\Delta \Omega^{(2)}_0) \equiv \frac{\big\vert \Delta \Omega^{(2)}_0(\text{THC}) - \Delta \Omega^{(2)}_0\big \vert}{\big \vert \Delta \Omega^{(2)}_0 \big \vert} \, ,
\end{align}
which goes to zero in the limit of an exact THC decomposition. Because of the highly-accurate Laplace transformed introduced above, the error presently propagates entirely from the approximation made on $^J{\tilde{\Omega}^{04}}$.

The right panel of Fig.~\ref{fig:bmbpt2} displays $\Delta (\Delta \Omega^{(2)}_0)$ for $^{18,20}$O as a function of $\epsilon$. A global trend is visible such that lower values of $\epsilon$ yield lower $\Delta (\Delta \Omega^{(2)}_0)$. One observes that the behavior is identical in both nuclei. Eventually, a THC approximation error of $\epsilon \approx 10^{-1}$ is sufficient to obtain $\Delta (\Delta \Omega^{(2)}_0)$ below $1\%$. Therefore, even though the matrix elements are only approximated to an accuracy of $10^{-1}$, the precision on the observable of interest is more than one order of magnitude better, which is achieved by retaining one order of magnitude less entries than in the original tensor. Even a quite crude approximation on the matrix elements thus yields an accuracy that is good enough to perform precision studies.

\section{Importance truncation}
\label{sec:it}

As an alternative to discarding high-rank components of the many-body tensors via a numerical factorization, a procedure to remove entries on the basis of an importance measure is now investigated.

\subsection{Basic concept}

The general idea is to discard irrelevant entries of the largest mode-$n$ tensors at play in the many-body framework of interest. This is done by a priori estimating the importance of each of its entries on the basis of a less costly many-body method than the envisioned one. In order to illustrate the concept, it is necessary to slightly reformulate the BCC method introduced in Sec.~\ref{Sec.BCC}. 

One considers so-called $P$ and $Q$ subspaces ${\cal F}^{(P)}$ and ${\cal F}^{(Q)}$ of Fock space ${\cal F}$ spanned by two selected sets of quasi-particle excitations of the Bogoliubov vacuum $| \Phi \rangle$ such that ${\cal F}^{(Q)} \subseteq ({\cal F}^{(0)}+{\cal F}^{(P)})^{\perp}$, where ${\cal F}^{(0)}$ is the one-dimensional subspace spanned by $| \Phi \rangle$. Typically, ${\cal F}^{(P)}$ is spanned by a selected set of low-rank excitations $\{| \Phi^{K} \rangle\}$ of $| \Phi \rangle$. Considering the cluster operator ${\cal T}^{(P)}$ associated with those excitations and solving the set of P-space BCC amplitude equations
\begin{equation}
\label{e:bcceqPQ}
{\cal M}_{K}(P) \equiv \langle \Phi^{K} | \bar{\Omega}^{(P)} | \Phi \rangle  =0 \, ,
\end{equation}
where $\bar{\Omega}^{(P)} \equiv e^{-\mathcal{T}^{(P)}} \Omega e^{\mathcal{T}^{(P)}}$, one obtains a first approximation to the ground-state energy via 
\begin{equation}
\label{e:bccenerPQfirst}
{\cal E}_{0}(P) \equiv \langle \Phi | \bar{\Omega}^{(P)} | \Phi \rangle \, .
\end{equation}
Once this is done, a non-iterative correction associated with quasi-particle excitations in ${\cal F}^{(Q)}$ is computed such that the energy eventually reads a
\begin{equation}
\label{e:bccenerPQ}
{\cal E}_{0} =  {\cal E}_{0}(P) + \delta(P;Q) \, .
\end{equation}
Several variants of (B)CC theories fit into the above formulation. For instance, setting ${\cal T}^{(P)}\equiv {\cal T}_{1}+\ldots {\cal T}_{r}$ and taking a null $Q$ space, standard BCC truncations with $\delta(P;Q)=0$ are recovered. Choosing the same $P$ space but computing $\delta(P;Q)$ in larger and larger $Q$ spaces via the method of moments, the Bogoliubov extension of the completely renormalized CC hierarchy~\cite{piecuch05a,piecuch06a,shen12a} is obtained. Using a more flexible definition of ${\cal F}^{(P)}$ while still computing $\delta(P;Q)$ via the method of moments, the Bogoliubov extension of the CC(P;Q) hierarchy~\cite{shen12a,shen12b,shen12c,bauman17a} is additionally obtained.

In the present paper, we wish to exploit the above formulation in a spirit close to BCC(P;Q) except that the energy correction $\delta(P;Q)$ is meant to be computed in perturbation and not via the more advanced method of moments\footnote{This choice is made for simplicity and using the method of moments is envisioned for future applications.}.

\subsection{Importance measure and ${\cal F}^{(P)}$ selection}

The general notion of IT has been developed in many different forms in the past. It already concerns the very initial truncation of the many-body basis or of the one-body basis used to represent the many-body tensors at play. 

In a conventional NCSM framework for example, one builds the $A$-body configuration space by including all particle-hole excitations below $\Xm{N}\hbar \Omega$, where $\Xm{N}$ denotes the sum of individual harmonic-oscillator excitation quanta with respect to the reference configuration. It corresponds to using $A$-body unperturbed excitation energies as an importance measure for the $A$-body basis states. 

In the present context, a similar truncation is at play when originally representing the $k$-body (mode-$2k$) operators (tensors) in the HO basis while employing a so-called "$e_{k\text{max}}$ truncation" to select the entries on the basis of the associated $k$-body HO energies. While one originally envisions to use a consistent scheme\footnote{This consistency is necessary for the truncated object initially representing the $k$-body operator to be an actual mode-$2k$ tensor.} for all tensors involved by using $e_{k\text{max}} = k e_{\text{max}}$, further truncations are typically envisioned for the largest $k$-body operators by relaxing the consistency between the various $k$-body sectors through the actual use of $e_{k\text{max}} < k e_{\text{max}}$. 

Given the initial representation of the tensors, the key point relates to the further selection of ${\cal F}^{(P)}$ and ${\cal F}^{(Q)}$. Typically, the goal is to approximate a full BCCSD or BCCSDT calculation at a (much) reduced computational cost. As such, one chooses ${\cal F}^{(P)}+{\cal F}^{(Q)}={\cal F}^{\text{SD}}$ or ${\cal F}^{\text{SDT}}$ and exploit the flexibility in the partitioning between $P$ and $Q$ to only solve the non-iterative equations in an optimal subspace ${\cal F}^{(P)}$ given a targeted accuracy. Aiming at BCCSD (BCCSDT), this translates into the fact that only a small subset of the original $t^{40}_{k_1 k_2 k_3 k_4}$ ($t^{60}_{k_1 k_2 k_3 k_4 k_5 k_6}$) tensor entries\footnote{Given that treating $\mathcal{T}_1$ in full is doable even for large $N$, single excitations are included in ${\cal F}^{(P)}$ by default.} is carried along when solving Eq.~\eqref{e:bcceqPQ} iteratively.

To actually select ${\cal F}^{(P)}$, i.e. the subset of $t^{40}_{k_1 k_2 k_3 k_4}$ ($t^{60}_{k_1 k_2 k_3 k_4 k_5 k_6}$) entries to be solved for, the simplest idea is to rely on a zeroth-order-like measure
\begin{align}
\kappa^{(0)} (t^{40}_{k_1 k_2 k_3 k_4\ldots}) \equiv E_{k_1 k_2 k_3 k_4 \ldots}^{-1} \, , \label{measure0}
\end{align}
which corresponds to selecting the tensor entries according to the sum of unperturbed quasi-particle energies associated with its $2k$ indices. While computationally inexpensive, such a scheme does neither take into account properties of the full Hamiltonian. Anticipating the too crude character of this truncation, one is led to evaluating the importance of the entries on the basis of a more optimal measure. Generically speaking, the relevance of a given importance measure relies on the following considerations
\begin{enumerate}
\item the measure must provide a robust estimate of the importance of the tensor entries delivered by the targeted calculation,
\item the evaluation of the importance measure must be as computationally inexpensive as possible, i.e. it must offer a significant gain compared to the evaluation of the tensor entries in the targeted calculation.
\end{enumerate}
In practice, a tradeoff between both requirements must be found. While the most precise estimate is obtained by evaluating the tensors of interest from a large-scale full configuration interaction calculation, it is of no practical interest since it already requires the full solution of the quantum many-body body problem. On the other hand a too simplistic estimate, e.g., through $\kappa^{(0)}$ introduced in Eq.~\eqref{measure0}, might not be related strongly enough to the full-fledged solution of interest.

In the present work, the lowest-order BMBPT counterpart of the tensors of interest is used as an importance measure. Alternatives, e.g. the early stages of FCIQMC or CCMC propagations~\cite{deustua17a}, can be envisioned if necessary. The first-order BMBPT estimates $\kappa^{(1)}$ of the double and triple amplitudes based a HFB vacuum and the NO2B approximation read as
\begin{subequations}
\label{TinBMBPT1st}
\begin{align}
t_{k_1 k_2 k_3 k_4}^{40(1)} &= -\frac{\Omega^{40}_{k_1 k_2 k_3 k_4} }{E_{k_1 k_2 k_3 k_4}} \, , \label{TinBMBPT1st2} \, \\
t_{k_1 k_2 k_3 k_4 k_5 k_6}^{60(1)} &= 0 \label{TinBMBPT1st3} \, .
\end{align}
\end{subequations}
The merit of $\kappa^{(1)}(t^{40}_{k_1 k_2 k_3 k_4})$ over $\kappa^{(0)}(t^{40}_{k_1 k_2 k_3 k_4})$ will be exemplified in realistic calculations below. 

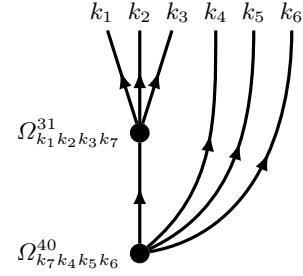
\begin{figure}
\begin{center}
\begin{tikzpicture}
\renewcommand{\vertexdistance}{2cm}
\node (1) at (0.0,0.8*\vertexdistance) {$k_2$};
\node[draw,omeganode,label=left:$\Omega^{31}_{k_1 k_2 k_3 k_7}$] (3) at (0,0) {};
\node[draw,omeganode,label=left:$\Omega^{40}_{k_7 k_4 k_5 k_6}$] (2) at (0,-0.8*\vertexdistance) {};
\node (4a) at (-0.5,0.8*\vertexdistance) {$k_1$};
\node (4b) at (0.5,0.8*\vertexdistance) {$k_3$};
\node (4c) at (1.0,0.8*\vertexdistance) {$k_4$};
\node (4d) at (1.5,0.8*\vertexdistance) {$k_5$};
\node (4e) at (2.0,0.8*\vertexdistance) {$k_6$};
\draw (3) edge[particle] node[left] {} (1);
\draw (2) edge[particle] node[left] {} (3);
\draw (2) edge[particle,out=40,in=-90] node[right] {} (4c);
\draw (2) edge[particle,out=25,in=-90] node[right] {} (4d);
\draw (2) edge[particle,out=10,in=-90] node[right] {} (4e);
\draw (3) edge[particle] node[right] {} (4a);
\draw (3) edge[particle] node[right] {} (4b);
\end{tikzpicture}
\end{center}
\caption{Hugenholtz diagram for the second-order BMBPT contribution to the triple BCC amplitude.}
\label{fig:wvdiagT3}
\end{figure}

While first-order BMBPT offers a $N^4$, i.e. a low cost, estimate of the double amplitudes, it does not do so for the triple amplitudes that are strictly zero at that order\footnote{In presence of $\Omega^{[6]}$, the leading triple amplitudes do arise at first order with a $N^6$ complexity.}. A non-zero estimate of the triple amplitudes requires to go to second order and reads as 
\begin{align}
t_{k_1 k_2 k_3 k_4 k_5 k_6}^{60(2)} &= +P(k_1k_2k_3/k_4k_5k_6)  \nonumber \\
& \hspace{0.4cm} \times \sum_{k_7} \frac{\Omega^{31}_{k_1 k_2 k_3 k_7}\, \Omega^{40}_{k_7 k_4 k_5 k_6}}{E_{k_7 k_4 k_5 k_6}E_{k_1 k_2 k_3 k_4 k_5 k_6}} \, , \label{TinBMBPT2nd3}
\end{align}
where the operator $P(k_1k_2k_3/k_4k_5k_6)$ permutes members of the triplet $(k_1,k_2,k_3)$ with those of the triplet $(k_4,k_5,k_6)$, i.e. it generates 20 different terms. The Hugenholtz diagram corresponding to Eq.~\eqref{TinBMBPT2nd3} is displayed in Fig.~\ref{fig:wvdiagT3}. Its evaluation is a costly non-iterative $N^7$ process. While constituting a state-of-the-art many-body calculation in itself, the evaluation of $t_{k_1 k_2 k_3 k_4 k_5 k_6}^{60(2)}$ remains very advantageous compared to the even more challenging iterative $N^8$ BCCSDT calculation of the $\mathcal{T}_3$ tensor entries. Full-fledged $J$-scheme calculations of $t_{k_1 k_2 k_3 k_4 k_5 k_6}^{60(2)}$ in large model spaces will be presented below to both set up the IT of a future BCCSDT calculation and to evaluate the corresponding $\delta(P;Q)$ perturbative correction to the energy. Performing the angular momentum recoupling of the mode-6 tensor $t_{k_1 k_2 k_3 k_4 k_5 k_6}^{60(2)}$ is also challenging in itself, especially because the recoupling is different for the 20 terms generated by $P(k_1k_2k_3/k_4k_5k_6)$ and spans several levels of complexity/computational cost. See App.~\ref{mode6coupling} for a detailed discussion of this point leading to the introduction of the $J$-coupled tensor 
\begin{equation}
{^{J_{12} J_{45} J} \tilde{t}}^{60}_{\tilde k_1 \tilde k_2 \tilde k_3 \tilde k_4 \tilde k_5 \tilde k_6} \equiv  \la [\tilde k_1 \tilde k_2 (J_{12})] \tilde k_3 (J) | {\cal T}_3 | [\tilde k_4 \tilde  k_5 (J_{45})] \tilde  k_6 (J) \ra \, , \nonumber
\end{equation}
whose second-order BMBPT estimate will be investigated below.

\begin{figure*}[t!]
\centering
\includegraphics{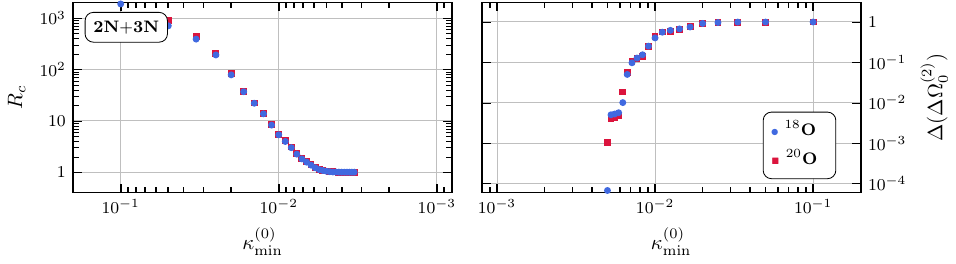}
\caption{Compression ratio and error on the second-order BMBPT energy correction in $^{18,20}$O as a function of the importance measure $\kappa^{(0)}_\text{min}$. Original tensors are built using an $\Xm{e}=4$ one-body HO basis. }
\label{fig:kap0}
\end{figure*}

\subsection{Importance-truncated tensor}

Considering the $n$-tuple BCC amplitude tensor
\begin{align}
\mathcal{T}_n \equiv \{t^{2n0}_{k_1 \ldots k_{2n}}\} \, ,
\end{align}
the corresponding \emph{importance-truncated} tensor based on the IT measure $\kappa^{(p)}(t^{2n0}_{k_1 \ldots k_{2n}})$ is obtained as
\begin{align}
\mathcal{T}_n(\kappa^{(p)}_\text{min}) \equiv \{t^{2n0}_{k_1 \ldots k_{2n}}  \,\, \text{such that}  \,\,  |t^{2n0(p)}_{k_1 \ldots k_{2n}}| \geq \kappa^{(p)}_\text{min} \} \, ,
\end{align}
where $\kappa^{(p)}_\text{min}$ defines the IT threshold. The original tensor is obtained in the limit of $\kappa^{(p)}_\text{min} \rightarrow 0$, i.e.,
\begin{align}
\lim_{\Xmin{\kappa^{(p)}}\rightarrow 0} \mathcal{T}_n(\kappa^{(p)}_\text{min}) = \mathcal{T}_n \, .
\end{align}
Additionally, for $\Xmin{\kappa^{(p)}} < {\Xmin{\kappa^{(p)\prime}}}$ the subset relation holds
 \begin{align}
\mathcal{T}_n(\kappa^{(p)\prime}_\text{min}) \subset \mathcal{T}_n(\kappa^{(p)}_\text{min}) \, ,
\end{align}
such that lowering $\kappa^{(p)}_\text{min}$ increases monotonically the number of entries in the truncated tensor.

\begin{figure*}[t!]
\centering
\includegraphics{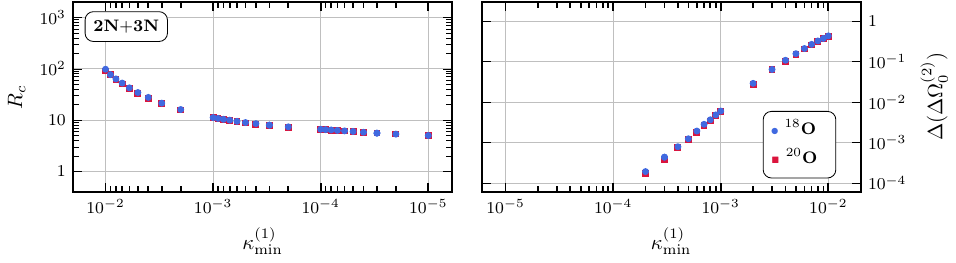}
\caption{Compression ratio and error on the second-order BMBPT energy correction in $^{18,20}$O as a function of the importance measure $\kappa^{(1)}_\text{min}$. Original tensors are built using an $\Xm{e}=4$ one-body HO basis.}
\label{fig:kap1}
\end{figure*}

\subsection{Small-scale IT applications}
\label{ITsmallscale}

The IT concept is first applied to the double BCC amplitude tensor $\mathcal{T}_2$ calculated for $^{18,20}$O in an $\Xm{e}=4$ one-body HO basis, i.e. for $N=140$ and $\tilde{N}=30$. This choice authorizes a direct comparison to the results previously obtained with TF techniques. 

The simplest importance measure $\kappa^{(0)}$ is investigated first, with identical results for both nuclei. The left panel of Fig.~\ref{fig:kap0} displays the compression ratio as a function of $\kappa^{(0)}_\text{min}$. For $\kappa^{(0)}_\text{min} \leq 5\cdot 10^{-2}$ the compression ratio is $R_c=1$ such that the IT-truncation is inactive. As expected, the compression ratio decreases monotonically when lowering the IT threshold. Reaching, e.g., a compression ratio $R_c=10$ corresponds to setting $\kappa^{(0)}_\text{min}\approx 10^{-2}$ and relates to an error $\Delta \mathcal{T}_2 = 0.5$ on the tensor. For comparison, the same compression ratio induced an error $\epsilon = 10^{-1}$ on the Hamiltonian tensor $\Omega^{[4]}$ in the THC framework. The right panel of Fig.~\ref{fig:kap0} displays the relative error on the second-order BMBPT correlation energy. For $R_c=10$, i.e. $\kappa^{(0)}_\text{min}\approx 10^{-2}$, a 50$\%$ error is generated, which compares very negatively to the sub-percent error obtained in the THC for a similar compression factor of the grand potential tensor.

\begin{figure}[t!]
\centering
\includegraphics{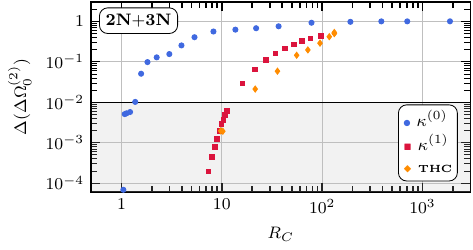}
\caption{Relative error $\Delta (\Delta \Omega^{(2)}_0)$ on the second-order BMBPT ground-state correlation energy against data compression factor $R_C$. Calculations are performed for $^{18}$O with initial tensors built in an $\Xm{e}=4$ one-body HO basis. Results are displayed for both THC and IT data compression techniques. In the latter case, both $\kappa^{(0)}$ or $\kappa^{(1)}$ are used as an important measure. The grey area indicates the region of sub-percent accuracy on $\Delta (\Delta \Omega^{(2)}_0)$.}
\label{fig:correl}
\end{figure}

As a next step, the left panel of Fig.~\ref{fig:kap1} displays $R_c$ as a function of $\kappa^{(1)}_\text{min}$ for the same nuclei. Again, the compression ratio decreases monotonically when lowering the IT threshold. For $\kappa^{(1)}_\text{min}=10^{-3}$, a compression ratio $R_c \approx 10$ is obtained for both nuclei, which corresponds to an error $\Delta \mathcal{T}_2 = 10^{-2}$ on the tensor. The associated relative error on the second-order BMBPT correlation energy (right panel) is equal to $1\%$. Lowering the IT threshold to $\kappa^{(1)}_\text{min}=3 \cdot10^{-4}$ barely reduces $R_c$ while decreasing the error by another order of magnitude.

Figure~\ref{fig:correl} summarizes the results by correlating $\Delta (\Delta \Omega^{(2)}_0)$ with $R_c$ for both TF and IT pre-processing methods. While operating in a very different fashion, TF based on THC and IT based on the first-order BMBPT estimation of $\mathcal{T}_2$ provide very similar performances. In particular, maintaining the relative error below $1\%$ can be achieved while compressing the data by one order of magnitude. While already satisfactory, Sec.~\ref{Sec.largescale} below illustrates how this result extends very favourably to larger model spaces and/or mode-6 tensors. It is also clear from Fig.~\ref{fig:correl} that employing a too naive IT measure, e.g. $\kappa^{(0)}$ based on unperturbed excitation energies of the associated configurations, is highly inefficient and must be discarded.

\subsection{Large-scale IT applications}
\label{Sec.largescale}

The conceptual simplicity of IT allows its  implementation and application in large model spaces. Furthermore, at the price of a state-of-the-art many-body development, the IT pre-processing is also applied to the mode-6 triple BCC amplitude tensor ${\cal T}_3$ in large model spaces.

\subsubsection{Double BCC amplitudes}

\begin{figure}[t!]
\centering
\includegraphics{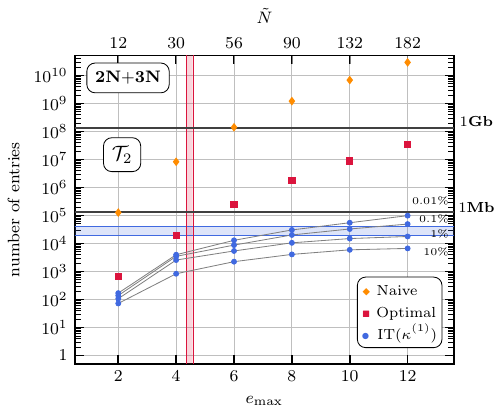}
\caption{Number of entries of $\mathcal{T}_2$ in $J$-scheme as a function  of $e_{\text{max}}$ (index range $\tilde{N}$): naive storage (yellow diamonds), symmetry optimized storage (red squares) and storage after IT truncation (blue circles) based on $\kappa^{(1)}$. For the latter, results are provided for several values of the relative error on the second-order BMBPT correlation energy in $^{18}$O. For convenience, the $1\,$Mb and $1\,$Gb storage limits in double precision are indicated. The horizontal blue band characterizes the number of entries in the IT tensor for $\Delta(\Delta \Omega_0^{(2)}) \in [0.1,1]\%$ using $e_{\text{max}}=12$. Intersecting it with the curve associated with the optimal storage scheme of the original $J$-coupled $\mathcal{T}_2$ tensor, one obtains the vertical red band defining an effective one-body basis size $\Xm{e}^\prime \in [4,5]$.}
\label{fig:ITdimT2}
\end{figure}

Applying IT to $\mathcal{T}_2$, the goal is to solve BCCSD equations for the retained entries (defining ${\cal F}^{(P)}$) and correct for the omitted ones (defining ${\cal F}^{(Q)}$) in perturbation. While solving CCSD in large model spaces is doable in both $J$- and $m$-scheme, BCCSD is more demanding and actually challenging given that the $N^{4}$ storage cost invokes the full basis dimension $N$ that cannot be separated into particle and hole states. In this context, IT can either provide a significant leverage in $J$-scheme\footnote{As seen below, the optimal storage cost of $\mathcal{T}_2$ in $J$-scheme is of the order of 200\,Mb for $e_{\text{max}}=12$.} or a mandatory leverage in $m$-scheme at the BCCSD level\footnote{The storage cost of $\mathcal{T}_2$ in $m$-scheme is of the order of 3\,Tb for $e_{\text{max}}=12$~\cite{Signoracci:2014dia}.}. While this will actually be investigated in a future publication, the present analysis is limited to characterizing the compression ratio obtained in $J$-scheme as a function of the IT threshold and to evaluating the error induced on the second-order BMBPT correlation energy by straightforwardly omitting the entries below the IT threshold.

The performance of IT techniques applied to the double BCC amplitude tensor $\mathcal{T}_2$ in a small model space has been characterized in Sec.~\ref{ITsmallscale}. The calculations are now extended up to $\Xm{e}=12$, i.e. up to $N=1820$ and $\tilde{N}=182$. In order to appreciate the data compression achieved for a given accuracy, Fig.~\ref{fig:ITdimT2} displays the number of entries of the $J$-coupled $\mathcal{T}_2$ tensor as a function of the size of the one-body HO basis. This is done for the naive and optimal storage schemes of the initial tensor as well as for the IT tensor based on $\kappa^{(1)}$. The numbers displayed for the initial tensor correspond to those given in Tab.~\ref{tabledimensions}. As for the IT tensor, results are provided for several values of the relative error on the second-order BMBPT correlation energy in $^{18}$O, ranging arbitrarily from $0.01\%$ to $10\%$. 

The first observation is that the naive and optimal storage schemes both exhibit exponential growth, however, with different rates. The naive scheme grows much more rapidly, finally requiring about three orders of magnitude more storage than the optimal one in $\Xm{e}=12$. The final storage requirement for the optimal scheme is about $0.5\,$Gb. While applying IT permitted to reduce the number of entries by one order of magnitude in an $\Xm{e}=4$ model space for a $1\%$ error on the correlation energy, the performance increases tremendously as a function of $\Xm{e}$. For $\Xm{e}=12$, the data is compressed by three orders of magnitude, six orders of magnitude compared to the naive storage scheme, for the same $1\%$ error on the correlation energy. Even for a very accurate $0.01\%$ error, one still obtains a data compression by more than two orders of magnitude. 

\begin{figure}[t!]
\centering
\includegraphics{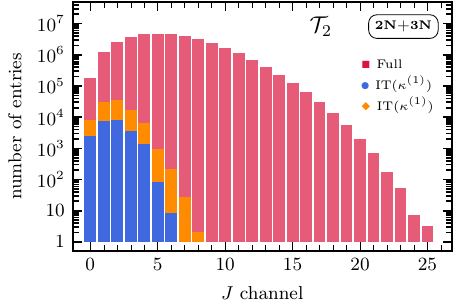}
\caption{(Color online) Distribution of entries of the $J$-coupled double BCC amplitude $\mathcal{T}_2$ as a function of the two-body angular-momentum $J$. Results in $^{18}$O are displayed before (red) and after IT truncation employing two different values of the important measure, i.e. $\kappa^{(1)}_{\text{min}}=9\cdot10^{-3}$ (orange) and $\kappa^{(1)}_{\text{min}}=10^{-5}$ (blue). Calculations are performed starting from an $\Xm{e}=12$ model space. }
\label{fig:compressionJ}
\end{figure}

In order to better characterize the effect of IT, Fig.~\ref{fig:compressionJ} compares the number of entries per $J$ block before and after IT. The IT results are displayed for $\kappa^{(1)}_{\text{min}}=9\cdot 10^{-3}$ and $\kappa^{(1)}_{\text{min}}=10^{-5}$, which correspond to $1\%$ and $0.01\%$ relative errors on the second-order energy correction, respectively. While a significant compression is observed for all $J$ values, a clear trend emerges in average: the larger the angular momentum associated to the entries, the lesser their importance. While the initial entries extend up to $J=2\Xm{e}+1=25$, there remains no entry beyond $J=8$ in the very precise IT tensor corresponding to $\kappa^{(1)}_{\text{min}}=10^{-5}$. No entries beyond $J=6$ are necessary to reach a $1\%$ error on the second-order energy correction.

In addition to the storage of $\mathcal{T}_2$, the CPU runtime is the other critical component of a computational analysis. While being computationally very simple, the evaluation of the second-order energy correction in large model spaces provides a useful testground for a CPU analysis. In order to produce a transparent comparison, the parallelization of the code was turned off for the present analysis.

\begin{figure}[t!]
\centering
\includegraphics{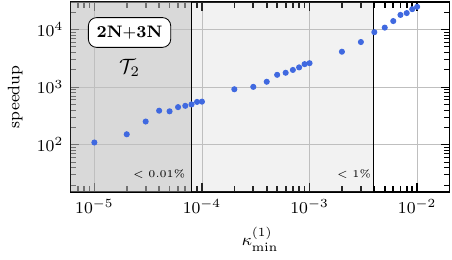}
\caption{Computational speedup as a function of the IT measure $\kappa^{(1)}_{\text{min}}$ for $^{18}$O. All calculations are performed starting from an $\Xm{e}=12$ model space. The limits associated to a relative error smaller than $1\%$ and $0.01\%$ on the second-order energy correction are indicated.}
\label{fig:runtime}
\end{figure}

Figure~\ref{fig:runtime} displays the speedup obtained by IT-BMBPT compared to the exact evaluation of the second-order correction in $^{18}$O. Two orders of magnitude in runtime are gained for very accurate calculations corresponding to $\kappa^{(1)}_{\text{min}} = 10^{-5}$ and four orders of magnitude are gained at the $1\%$ error level. Whereas the goal is not to obtain a speedup for a low-cost theory such as BMBPT(2), these numbers are very encouraging in view of performing IT-BCCSD calculations in the future.

\begin{figure}[t!]
\centering
\includegraphics{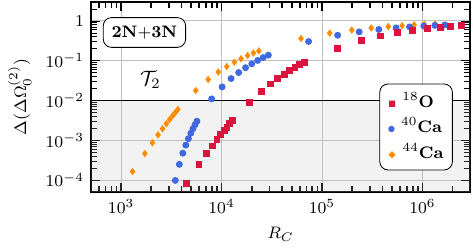}
\caption{Relative error $\Delta (\Delta \Omega^{(2)}_0)$ on the second-order BMBPT ground-state correlation energy against IT data compression factor $R_C$. Results are displayed for $^{18}$O, $^{40}$Ca and $^{44}$Ca starting from an $\Xm{e}=12$ HO basis.}
\label{fig:pairing}
\end{figure}

\begin{figure}[t!]
\centering
\includegraphics[width=1.0\columnwidth]{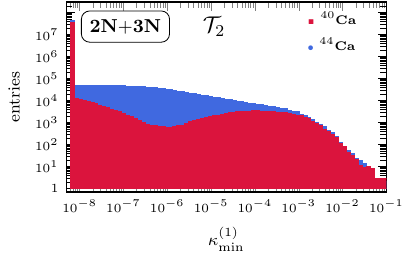}
\caption{Histogram of the importance measure $\kappa^{(1)}_{\text{min}}$ for $^{40}$Ca and $^{44}$Ca obtained in an $\Xm{e}=12$ model space.}
\label{fig:hist}
\end{figure}

\begin{figure}[t!]
\centering
\includegraphics[width=1.0\columnwidth]{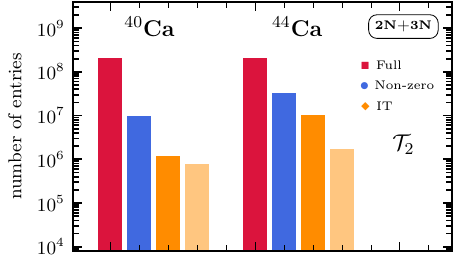}
\caption{ Histogram of the number of tensor entries of ${\cal T}_2$ in the optimal storage scheme (\redsquare), the number of non-zero entries (\bluecircle) and the number of entries of the associated IT tensors (\orangediamond) for $\kappa^{(1)}_{\text{min}} = 10^{-7}$ and $\kappa^{(1)}_{\text{min}} = 10^{-5}$.}
\label{fig:pairingsparsity}
\end{figure}

So far, results have been displayed for weakly-paired Oxygen isotopes\footnote{In prior BMBPT calculations using the same Hamiltonian, oxygen isotopes were identified to be only weakly paired as testified by the small particle-number variance~\cite{Tichai:2018mll}.}. We now wish to investigate how the IT technique behaves for heavier nuclei and when going from a closed-shell to an open-shell system . To do so, IT calculations are repeated for the unpaired doubly closed-shell $^{40}$Ca nucleus and for the singly open-shell $^{44}$Ca isotope that exhibits stronger pairing\footnote{Note however that $^{44}$Ca computed with the same Hamiltonian via GSCGF theory at the ADC(2) truncation level, which is  close to BMBPT(2)~\cite{Tichai:2018mll}, displays only about half of the experimental pairing gap measured via the three-point mass difference formula~\cite{Barb18}.} than $^{18,20}$O.

Figure~\ref{fig:pairing} displays the relative error on the second-order BMBPT ground-state correlation energy against the data compression factor for $^{18}$O and $^{40,44}$Ca. Comparing first the two paired systems $^{18}$O and $^{44}$Ca, one observes a mass dependence while working with a fixed basis size ($\Xm{e}=12$) that is relatively larger to begin with for $^{18}$O. While the two curves follow the same overall trend, the data compression achieved for a given accuracy is larger in $^{18}$O, e.g. the $1\%$ accuracy achieved for $R_c=2\cdot10^3$ in $^{18}$O is only reached for $R_c=4\cdot10^2$ in $^{44}$Ca. Focusing next on $^{40}$Ca and $^{44}$Ca, the curves follow qualitatively two different trends. A significantly smaller error is generated in $^{40}$Ca than in $^{44}$Ca when truncating the smallest entries. Because $^{40}$Ca is not superfluid, the associated HFB vacuum reduces to a HF Slater determinant such that the matrix elements of $^{J}{\tilde{\Omega}^{40}}$, and thus of $^{J}{\tilde{t}^{40(1)}}$, display a particle-hole symmetry and thus more zero entries than for $^{44}$Ca to begin with. It eventually leads to a more efficient data compression for $^{40}$Ca. These characteristics are visible in Fig.~\ref{fig:pairingsparsity}.  For example, the compression obtained for a $1\%$ error is twice as better in $^{40}$Ca than in $^{44}$Ca. For errors larger than $10\%$, the benefit associated with the particle-hole symmetry for high accuracy calculations disappears and both curves converge towards one another such that the closed- or open-shell character of the system becomes irrelevant.

\subsubsection{Triple BCC amplitudes}

Applying IT to $\mathcal{T}_3$, the goal is to solve BCCSDT equations for the retained entries (defining ${\cal F}^{(P)}$) and correct for the omitted ones (defining ${\cal F}^{(Q)}$) in perturbation. The feasibility of the approach directly depends on the reduction offered by the IT for a desired accuracy given that a full BCCSDT calculation is currently undoable in realistic model spaces, even in $J$-scheme. 

\begin{figure}[t!]
\centering
\includegraphics{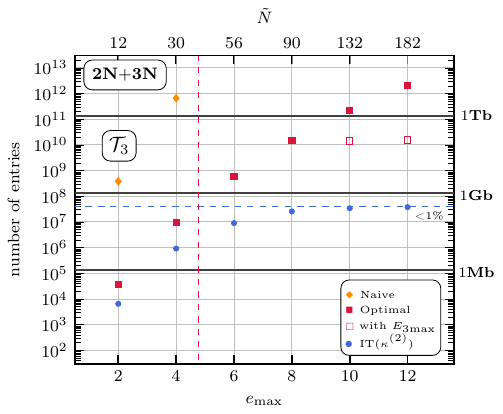}
\caption{Number of entries of $\mathcal{T}_3$ in $J$-scheme as a function of $e_{\text{max}}$ (index range $\tilde{N}$) for the naive (\orangediamond) and optimal storage (\redsquare) schemes. For $e_{\text{max}}=10,12$, further truncating entries to $e_{3\text{max}}=22<3\cdot e_{\text{max}}$ delivers the open red squares. The number of entries after IT based on $\kappa^{(2)}$ are also displayed (\bluecircle), starting from the open red squares for $e_{\text{max}}=10,12$. The IT results are provided for a $1\%$ relative error on the fourth-order BMBPT correlation energy from the leading triple BCC amplitude in $^{18}$O. For convenience, the $1\,$Mb, $1\,$Gb and $1\,$Tb storage limits in double precision are indicated. The horizontal dashed blue line characterizes the number of entries in the IT tensor for $\Delta(\Delta\Omega^{[4_T]}_0) =1\%$ using $e_{\text{max}}=12$.  Intersecting it with the curve associated with the optimal storage scheme of the original $J$-coupled $\mathcal{T}_3$ tensor, one obtains the vertical red line defining an effective one-body basis size $\Xm{e}^\prime \approx 5$.}
\label{fig:ITdimT3}
\end{figure}

The IT measure requires the perturbative evaluation of $\mathcal{T}_3$ as defined by Eq.~\eqref{TinBMBPT2nd3} (see Eq.~\eqref{eq:kappa2J} for the $J$-coupled form), which is a $N^7$ process. Based on this estimate, the associated contribution to the  fourth-order ground-state correlation energy
\begin{align}
\Delta \Omega^{[4_T]}_0 = \sum_{ \substack{k_1 k_2 k_3 \\ k_4 k_5 k_6} }  |t^{60 (2)}_{k_1 k_2 k_3 k_4 k_5 k_6} |^2 E_{k_1 k_2 k_3 k_4 k_5 k_6} \, ,
\label{eq:Etriplesbody}
\end{align}
which is necessarily positive, is used to gauge the propagated error. The evaluation of $\Delta \Omega^{[4_T]}_0$ is itself a $N^6$ process.

In order to appreciate the data compression achieved for a given accuracy, Fig.~\ref{fig:ITdimT3} displays the number of entries of the $J$-coupled $\mathcal{T}_3$ tensor as a function of the size of the one-body basis. This is done for the naive and optimal storage schemes of the initial tensor as well as for the IT tensor based on $\kappa^{(2)}$. The numbers displayed for the initial tensor correspond to those given in Tab.~\ref{tabledimensions}. For $e_{\text{max}}=10,12$, additional numbers are given for the optimal storage scheme that correspond to performing an additional reduction to $e_{3\text{max}}=22<3\cdot e_{\text{max}}$. This additional truncation, which still leads to working with an extremely large mode-6 tensor, is necessary to evaluate $t_{k_1 k_2 k_3 k_4 k_5 k_6}^{60(2)}$. The IT results correspond to a $1\%$ relative error on the second-order BMBPT correlation energy in $^{18}$O and are based on the initially reduced tensor for $e_{\text{max}}=10,12$.

Once again, the naive and optimal storage schemes both exhibit exponential growth. The storage requirements in the optimal scheme are about $15\,$Tb in $\Xm{e}=12$ for $e_{3\text{max}}=36$ and $200\,$Gb for the reduced $e_{3\text{max}}=22$, which is  unmanageable in a BCCSDT calculation. While the IT permits to reduce the number of entries by one order of magnitude at $\Xm{e}=4$ for a $1\%$ error on $\Delta \Omega^{[4_T]}_0$, the performance increases tremendously as a function of $\Xm{e}$. For $\Xm{e}=12$ and $e_{3\text{max}}=22$, the data is compressed by 2 orders of magnitude\footnote{Starting from $e_{3\text{max}}=36$, the compression factor would undoubtedly be significantly larger than 2 orders of magnitude such that the number of remaining entries would probably not be much larger than in the present calculation.} for the same $1\%$ error such that the IT tensor to handle requires less than $1\,$Gb. As illustrated in Tab.~\ref{DeltaOmega0}, it must be noted that $\Delta \Omega^{[4_T]}_0$ being one order of magnitude smaller than $\Delta \Omega^{[2]}_0$, requiring $1\%$ error on the former is equivalent to requiring $0.1\%$ error on the latter, i.e. an error of $30$\,keV out of a total binding energy of about $135$\,MeV in $^{18}$O. This level of error is largely negligible given the predictive power of current \emph{ab initio} calculations.

\begin{table}[t!]
\def\arraystretch{1.6}
\centering
\begin{tabular}{c  | c | c | c}
\hline \hline 
$e_\text{max}$  &   $E_\text{HFB}$   $\left [\text{MeV}\right ]$ & $\Delta \Omega_0^{(2)}$ $\left [\text{MeV}\right ]$ & $\Delta \Omega^{[4_\text{T}]}_0$  $\left [\text{MeV}\right ]$  \\
\hline \hline
2  & -100.481 & -10.328 & 1.228 \\
4  & -105.994 & -25.815 & 2.493 \\
6  & -107.289$^{\dagger}$ & -29.316$^{\dagger}$ & 2.961$^{\dagger}$ \\
8  & -107.588$^{\dagger}$ & -29.834$^{\dagger}$ & 3.022$^{\dagger}$ \\
10  & -107.802$^{\dagger}$ & -29.875$^{\dagger}$ & 2.962$^{\dagger\ast}$ \\
12  & -108.051$^{\dagger}$ & -29.783$^{\dagger}$ & 2.956$^{\dagger\ast}$ \\
\hline \hline
\end{tabular}
\caption{BMBPT contributions to the binding energy of $^{18}$O as a function of $e_\text{max}$; i.e. first-order (HFB) contribution, second-order correction and fourth-order correction from the leading triple BCC amplitude. Because the original matrix elements of the three-nucleon force in Eq.~\eqref{eq:ham} are limited by an a priori $e_{3\text{max}}=14$ truncation, numbers marked by $^{\dagger}$ are slightly approximate compared to the calculation that would consistently use $e_{3\text{max}}=3\,e_{\text{max}}$. Furthermore, numbers marked by $^{\ast}$ have been obtained by imposing an initial $e_{3\text{max}}=22$ reduction on the leading triple BCC amplitude tensor.}
\label{DeltaOmega0}
\end{table}

\begin{figure}[t!]
\centering
\includegraphics{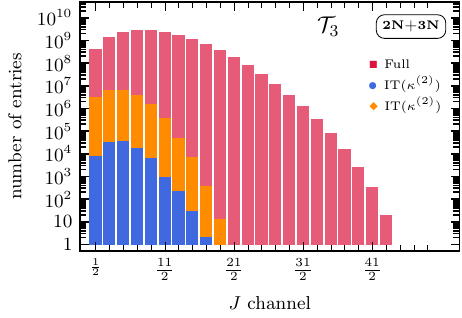}
\caption{(Color online) Distribution of entries of the $J$-coupled triple BCC amplitude $\mathcal{T}_3$ as a function of the three-body angular-momentum $J$. Results in $^{18}$O are displayed before (red) and after IT truncation employing two different values of the important measure, i.e. $\kappa^{(2)}_{\text{min}}=10^{-6}$ (orange) and $\kappa^{(2)}_{\text{min}}=6\cdot 10^{-6}$ (blue). Calculations are performed using $\Xm{e}=12$ and $e_{3\text{max}}=22$. }
\label{fig:compressionJ3}
\end{figure}

In order to better characterize the effect of IT, Fig.~\ref{fig:compressionJ3} compares the number of entries per $J$ block before and after IT. The IT results are displayed for $\kappa^{(2)}_{\text{min}}=10^{-6}$ and $\kappa^{(2)}_{\text{min}}=6\cdot10^{-6}$, which corresponds to $1\%$ and $5\%$ relative error on $\Delta \Omega^{[4_T]}_0$, respectively. While a significant compression is observed for all $J$ values, the same trend emerges as for $\mathcal{T}_3$, i.e. the larger the three-body angular momentum associated to the entries, the lesser their importance. While the initial entries extend up to $J=43/2$, there remains no entry beyond $J=19/2$ for $\kappa^{(2)}_{\text{min}}=10^{-6}$. No entries beyond $J=17/2$ are necessary to reach a $5 \%$ error on $\Delta \Omega^{[4_T]}_0$.

\subsection{Discussion}
\label{sec:discussion}

Let us now make three comments on IT to anticipate further developments
\begin{enumerate}
\item While the entire benefit of the IT cannot be obtained in this way, it does emerge a posteriori that a large part of its effects is to eliminate the entries entering large-$J$ blocks. A systematic study as a function of $\Xm{e}$ could allow us to understand quantitatively which two-body and three-body $J$ blocks are entirely irrelevant for a given accuracy on the correlation energy. Based on the results of such a study, one could avoid producing those $J$ blocks from the outset, i.e. design an a priori truncation with respect to $J$, thus leading to the need to run the IT pre-processing on much smaller tensors to begin with. This could help pushing calculations to yet higher-mode tensors, i.e. higher orders in the many-body expansion, and/or larger $\Xm{e}$.
\item The benefits of IT are expected to be even more pronounced when going to doubly open-shell nuclei that necessitate to work in an $m$-scheme basis. For $\Xm{e}=12$, a full-fledged BCCSD calculation would require over $3$ Tb of storage~\cite{Si15} such that only techniques like IT could make such a calculation manageable.
\item  Similar to many-body tensors in CC theory, IT can be straightforwardly applied to Bogoliubov configuration interaction truncated up to, e.g., 8 quasi-particle excitations yielding the IT-BCISDTQ approach. Doing so requires the derivation of the $\kappa^{(2)}$ estimate of the leading quadruple contributions to the BMBPT wave function. Applying Lanczos diagonalization, ground-state and low-lying excitation energies could be accessed, as well as other observables. Eventually, particle-number symmetry can be consistently restored by applying projection techniques. The lack of size-extensivity of truncated (B)CI calculations can be approximately cured via an \emph{a posteriori} corrections of Davidson type.
\end{enumerate}

\section{Computational analysis of TF and IT}

Typically, expansion methods admit polynomial scaling with respect to system size\footnote{In practice, the cost is also driven by a prefactor that can be sufficiently different in two methods to obscure the classification associated to the scaling law over a certain interval of $N$ values. }
\begin{align}
\mathcal{O}(N^\alpha) \, ,
\end{align}
where $\alpha$ is a characteristic exponent of the theory\footnote{Valence-space approaches involve the diagonalization of a dressed Hamiltonian in an active space of limited size. Consequently, such methods are excluded from the present analysis even though the dressing of the Hamiltonian itself requires only polynomial effort.}. Table~\ref{tab:scaling} displays the value of $\alpha$ for a collection of many-body methods routinely used in nuclear theory. Furthermore, the order of MBPT-completeness associated to each method/truncation is added, i.e., the order up to which all MBPT diagrams are included.  While TF and IT aim at reducing the computational effort required to solve the many-body problem, they operate differently to achieve this goal. 

\begin{table}
\center
\def\arraystretch{1.3}
\begin{tabular}{c | c | c}
Truncation scheme & Scaling & (B)MBPT(n) \\
\hline \hline
(B)MBPT(2) & $N^4$ & 2\\
(B)MBPT(3) & $N^6$ & 3\\
(B)MBPT(4) & $N^7$ & 4\\
 \hline\hline
(B)CCSD & $N^6$ & 3\\
(B)CCSDT & $N^8$ & 4\\
 \hline\hline
IMSRG(2) & $N^6$ &3\\
ADC(3) & $N^6$  & 3\\
 \hline\hline
 \end{tabular}
\caption{Computational scaling of state-of-the-art expansion many-body methods. The displayed numbers correspond to optimized contraction patterns obtained via the introduction of intermediates in MBPT and CC or via the use of the natural basis in IMSRG.}
\label{tab:scaling}
\end{table}

First, TF potentially allows one to decrease the exponent to an effective value $\alpha^\prime$ by decomposing many-body tensors and finally only invoking contractions between low-rank operators. Assuming a mild scaling of $\rk{THC}$ with $N$, the complexity of the tensor network might be drastically reduced, i.e. ideally, CC theory at arbitrary truncation level could be evaluated at $N^4$ if the THC ranks admit (nearly) linear scaling. Although it is unclear at this point if this formal counting is contaminated in practice by a large prefactor, it is indeed very promising. 

Second, while IT does not act on the scaling exponent, it decreases the size of the one-body basis to an effective value $N^\prime$ by only keeping a small subset of the entries. Focusing first on the double BCC amplitude ${\cal T}_2$, Fig.~\ref{fig:compressionJ} shows that for $\Xm{e}=12$ ($\tilde{N} \approx 182$), about $10^{5}$ tuples are retained after the IT truncation when targeting a very conservative $0.01\%$ error on the second-order correlation energy, which effectively corresponds to $\tilde{N}^\prime \approx 45$ in the optimal storage scheme, i.e. to an effective number of major HO shells $\Xm{e}^\prime \approx 5$. Moving to the triple BCC amplitude ${\cal T}_3$, Fig.~\ref{fig:ITdimT3} shows that less than $10^{8}$ tuples are retained after the IT truncation when targeting a conservative $1\%$ error on the fourth-order correlation energy. This yields a very similar effective one-body basis size $\tilde{N}^\prime \approx 45$, i.e. $\Xm{e}^\prime \approx 5$. In conclusion, although IT does not change the scaling of the theory, it allows one to perform high-accuracy non-perturbative calculations in a realistic model space characterized by, e.g., $\Xm{e}=12$ at an effective price corresponding to a much reduced model space characterized by $\Xm{e}^\prime \approx 5$. This is very promising.

\section{Conclusions}
\label{sec:conclusion}

In this work tensor-factorization and importance-truncation techniques are introduced as two different paradigms to pre-process the nuclear many-body problem with the goal to push ab initio calculations based on expansion methods to (i) higher accuracies, (ii) doubly open-shell nuclei and (iii) nuclei with $A>100$. 

Bogoliubov many-body perturbation theory calculations of semi-magic nuclei are used in small model spaces to test the numerical implementation of tensor-factorization and importance-truncation techniques. Both methods yield very promising computational advantages with respect to storage requirements of many-body tensors while generating a small error on nuclear ground-state energies. Eventually, large-scale IT benchmarks based on the state-of-the-art computation of perturbative triple amplitudes in Bogoliubov coupled cluster theory confirm the great promises of this pre-processing method. 

The next step consists of implementing THC and IT techniques in actual non-perturbative calculations, which seems particularly straightforward for IT. While these pre-processing method can be adapted to any method, our focus will be on Bogoliubov extensions of CC and truncated CI with the goal to address open-shell physics from simple single-reference methods.

\section*{Acknowledgements }

We thank Robert Roth for providing us with nuclear NN and 3N matrix elements. This publication is based on work supported in part by the framework of the Espace de Structure et de r\'eactions Nucl\'eaires Th\'eorique (ESNT) at CEA.

\begin{appendix}
\setcounter{equation}{0}
\renewcommand\theequation{A.\arabic{equation}}
\allowdisplaybreaks

\section{Angular momentum coupling}
\label{sec:AMC} 

In all applications performed in this work a spherical formulation of the underlying many-body theory is employed and, consequently, the evaluation of working expressions must be performed in an angular-momentum coupled scheme.
The aim of this section is to discuss the implementational details of the evaluation of importance measures in such a spherical scheme requiring an angular-momentum reduction of the underlying many-body diagrams. While the basic notations and coupling symbols are presently introduced, the reader is referred to Ref.~\cite{VaMo88} for an extensive treatment of angular-momentum theory.

\subsection{Basics}

A generic state of the one-body Hilbert space ${\cal H}_1$ is denoted as $|k\rangle$, where $k$ stands for a collective index
\begin{align}
k \equiv (n_k, l_k, j_k ,t_k , m_k)\, ,
\end{align}
with $n_k$ the radial quantum number, $l_k$ the orbital angular-momentum quantum number, $j_k$ the total angular-momentum quantum number with projection $m_k$ and  $t_k$ the isospin projection distinguishing protons and neutrons. A reduced set of quantum numbers is introduced through
\begin{align}
\tilde k \equiv (n_k, l_k, j_k ,t_k)\, ,
\label{eq:ktilde}
\end{align}
where the angular-momentum projection is explicitly excluded. Effectively, Eq.~\eqref{eq:ktilde} contains all quantum numbers necessary to describe a rotationally invariant system. Forming the tensor product of two one-body states, basis states of the two-body Hilbert space ${\cal H}_2$ are obtained in the uncoupled representation as
\begin{align}
| k_1 k_2 \rangle \equiv |k_1 \rangle \otimes |k_2 \rangle \, .
\end{align}
In the coupled representation, the two total angular momenta of the states $|k_1\rangle$ and $|k_2 \rangle$ are coupled to a total two-body angular momentum $J$ and projection\footnote{While the two-body state does indeed depend on $M$, the label is omitted for brevity given that the reduced tensors eventually built in that basis are diagonal in $M$ and independent of it.} $M$,
\begin{align}
| \tilde k_1 \tilde k_2 (J) \ra = \sum_{m_{k_1} m_{k_2}} \clebsch{j_{k_1}}{m_{k_1}}{j_{k_2}}{m_{k_2}}{J}{M} \, |k_1 k_2 \rangle \, ,
\end{align}
where $\clebsch{j_{k_1}}{m_{k_1}}{j_{k_2}}{m_{k_2}}{J}{M}$ denotes the Clebsch-Gordan coefficient (CGC) mitigating the transformation from the uncoupled to the coupled basis. The inverse transformation is given by
\begin{align}
|  k_1 k_2 \ra = \sum_{JM} \clebsch{j_{k_1}}{m_{k_1}}{j_{k_2}}{m_{k_2}}{J}{M} \, | \tilde k_1 \tilde k_2 (J) \rangle \, .
\end{align}
Applying the same rationale to the bra two-body states leads to defining the $J$-coupled representation of, e.g., the interaction matrix elements $^{J}v_{\tilde k_1 \tilde k_2 \tilde k_3 \tilde k_4}$ via
\begin{align}
v_{k_1 k_2 k_3 k_4} \equiv \sum_{JM} \clebsch{j_{k_1}}{m_{k_1}}{j_{k_2}}{m_{k_2}}{J}{M} \clebsch{j_{k_3}}{m_{k_3}}{j_{k_4}}{m_{k_4}}{J}{M} \, ^{J}v_{\tilde k_1 \tilde k_2 \tilde k_3 \tilde k_4}\, .
\end{align}
Due to rotational invariance, nuclear matrix elements are diagonal with respect to the two-body angular momenta of the bra and ket states and independent of their projection $M$.

Furthermore, angular-momentum coupling is extended to basis states of  ${\cal H}_3$, i.e., the tensor product of three single-particle states 
\begin{align}
| k_1 k_2 k_3 \rangle \equiv |k_1 \rangle \otimes |k_2 \rangle \otimes |k_3 \rangle \, .
\end{align}
In order to do so a coupling order needs to be fixed. In the subsequent derivations this is chosen to be
\begin{align}
| k_1 k_2 k_3 \rangle  = \sum_{ \substack{J_{12} J \\ M_{12} M }}
\clebsch{j_{k_1}}{m_{k_1}}{j_{k_2}}{m_{k_2}}{J_{12}}{M_{12}}
\clebsch{J_{12}}{M_{12}}{j_{k_3}}{m_{k_3}}{J}{M} \notag \\ \times | [\tilde k_1 \tilde k_2 (J_{12})] \tilde k_3 (J) \rangle \, ,
\end{align}
where the intermediate two-body angular-momentum quantum number $J_{12}$ is integer and the three-body angular-momentum quantum number $J$ is half-integer.

While the initial choice of the coupling order for three-body states is arbitrary, a consistent treatment throughout the derivation is crucial. A proper selection of this coupling order may significantly simplify or complicate the resulting final expressions in the many-body framework.

\subsection{Quasi-particle matrix elements}

Since the building blocks, e.g. grand potential matrix elements, of a particle-number broken many-body formalism are defined in quasi-particle space, the angular-momentum coupling needs to be extended to these general objects. However, the different normal-ordered components admit different symmetry properties with respect to parity and $M$ conservation. In order to unify the treatment the notion of \textit{cross-coupled} matrix elements is introduced yielding matrix elements with the same block structure for all $\Omega^{ij}$ components~\cite{Tichai19unp}. In the following all working equations are expressed in terms of such cross-coupled matrix elements indicated by a tilde on top of the corresponding symbols, e.g., $\tilde \Omega^{ij}$ instead of $\Omega^{ij}$.

The angular-momentum coupling of cross-coupled matrix elements of the normal-ordered grand-potential components reads as
\begin{subequations}
\begin{align}
\tilde{\Omega}^{40}_{k_1 k_2 k_3 k_4} &= \sum_{JM} (-1)^{j_{k_3} + j_{k_4} - m_{k_3} -m_{k_4}}\clebsch{j_{k_1}}{m_{k_1}}{j_{k_2}}{m_{k_2}}{J}{M} \nonumber \\ 
& \hspace{1.5cm} \times \clebsch{j_{k_3}}{m_{k_3}}{j_{k_4}}{m_{k_4}}{J}{M} {^J} \tilde{\Omega}^{40}_{\tilde k_1 \tilde k_2 \tilde k_3 \tilde k_4}\, , \\
\tilde{\Omega}^{31}_{k_1 k_2 k_3 k_4} &= \sum_{JM} (-1)^{j_{k_3}  -m_{k_3}} \clebsch{j_{k_1}}{m_{k_1}}{j_{k_2}}{m_{k_2}}{J}{M} \nonumber \\ 
&  \hspace{1.5cm} \times  \clebsch{j_{k_3}}{m_{k_3}}{j_{k_4}}{m_{k_4}}{J}{M} {^J} \tilde{\Omega}^{31}_{\tilde k_1 \tilde k_2 \tilde k_3 \tilde k_4}\, , \\
\tilde{\Omega}^{22}_{k_1 k_2 k_3 k_4} &= \sum_{JM} \clebsch{j_{k_1}}{m_{k_1}}{j_{k_2}}{m_{k_2}}{J}{M} \nonumber \\ 
&  \hspace{1.5cm} \times  \clebsch{j_{k_3}}{m_{k_3}}{j_{k_4}}{m_{k_4}}{J}{M} {^J} \tilde{\Omega}^{22}_{\tilde k_1 \tilde k_2 \tilde k_3 \tilde k_4}\, , 
\end{align}
\label{eq:omegaJ}
\end{subequations}
where the actual expression of angular-momentum coupled quantities ${^J}\tilde{\Omega}^{ij}$ will be detailed in a forthcoming publication. Note the appearance of an additional ($m$-dependent) phase factor for $\tilde{\Omega}^{40}$ and $\tilde{\Omega}^{31}$ compared to $\tilde{\Omega}^{22}$.

When performing the angular-momentum reduction of complex tensor networks the use of recoupling symbols is inevitable. Of particular importance is the \emph{Wigner 6j-symbol}
\begin{align}
\sixj{j_{k_1}}{j_{k_2}}{J}{j_{k_3}}{j_{k_4}}{J^\prime} \, ,
\end{align}
which arises naturally from the coupling of three angular momenta~\cite{VaMo88}. In principle higher recoupling symbols like $9j$- and $12j$-symbols may also appear. However, due to limited cache size, it is convenient to precompute and store only $6j$-symbols. Thanks to angular-momentum identities, higher-order recoupling symbols can be reexpressed as products of $6j$-symbols that indeed constitute indeed the most complex recoupling symbols needed in our calculations.

The reduction of both $\tilde t^{40(1)}$ and $\tilde t^{60(2)}$ is now performed. The construction of $\tilde t^{20 (1)}$ or $\tilde t^{20 (2)}$ can be done analogously. However, the low number of entries of $t^{20}$ typically does not require an IT treatment.

\subsection{Reduction of $\tilde t^{40(1)}$}

The cross-coupled matrix elements of ${\cal T}_2^{(1)}$ read in $m$-scheme as
\begin{align}
\tilde t_{k_1 k_2 k_3 k_4}^{40(1)} &= -\frac{\tilde \Omega^{40}_{k_1 k_2 k_3 k_4} }{E_{k_1 k_2 k_3 k_4}} \, , \label{eq:kappa1sph}
\end{align}
and is thus trivially proportional to $\tilde \Omega^{40}$ such that no explicit permutation of external indices needs to be applied to ensure the full antisymmetry of $\tilde t^{40(1)}$. Inserting $J$-coupled matrix elements of $\tilde \Omega^{40}$ in Eq.~\ref{eq:kappa1sph}, one straighforwardly obtains those of interest. While the $m$-scheme expression requires $N^4$ evaluations, its $J$-coupled partner \eqref{eq:kappa1sph} only necessitates $J_{2\text{max}} \cdot \tilde N^4$ evaluations where $J_{2\text{max}}$ defines the number of channels of the two-body angular-momentum $J$. The number of $J$ channels is $J_{2\text{max}} = 2 e_\text{max}+2$ as long as no additional truncation on $e_{2\text{max}} \equiv e_1 + e_2$ is employed.

\subsection{Reduction of $\tilde t^{60(2)}$}
\label{mode6coupling}

The $m$-scheme cross-coupled matrix elements $\tilde t^{60 (2)}_{k_1k_2 k_3 k_4 k_5 k_6}$ associated with the diagram
displayed in Fig.~\ref{fig:wvdiagT3} and whose algebraic form was given in Eq.~\eqref{TinBMBPT2nd3} can be written in a compact form as
\begin{align}
\tilde t^{60 (2)}_{k_1k_2 k_3 k_4 k_5 k_6} &= -P(k_1k_2k_3/k_4k_5k_6)  \nonumber \\
& \hspace{0.5cm} \times \sum_{k_7} \frac{\tilde \Omega^{31}_{k_1 k_2 k_3 k_7}\, \tilde t^{40(1)}_{k_7 k_4 k_5 k_6}}{E_{k_1 k_2 k_3 k_4 k_5 k_6}} \, ,
\label{eq:kappa2}
\end{align}
where the operator $\mathcal{P}(k_1 k_2 k_3 / k_4 k_5 k_6)$ permutes indices from the first group with indices from the second group in all possible ways. The permutation operator generates $20$ different terms yielding a complex expression for $\tilde t^{60 (2)}$. 

The angular-momentum coupling of the cross-coupled matrix elements of the triple BCC amplitude reads as
\begin{align}
{^{J_{12} J_{45} J}} \tilde t^{60(2)}_{\tilde k_1 \tilde k_2 \tilde k_3 \tilde k_4 \tilde k_5 \tilde k_6} =&\sum_{ \substack{m_{k_1} m_{k_2} m_{k_3} \\ m_{k_4} m_{k_5} m_{k_6} }} 
(-1)^{j_{k_4} + j_{k_5}+j_{k_6} - M} \notag \\
& \times \clebsch{j_{k_1}}{m_{k_1}}{j_{k_2}}{m_{k_2}}{J_{12}}{M_{12}}
\clebsch{J_{12}}{M_{12}}{j_{k_3}}{m_{k_3}}{J}{M} \notag \\ &\times
\clebsch{j_{k_4}}{m_{k_4}}{j_{k_5}}{m_{k_5}}{J_{45}}{M_{45}} 
\clebsch{J_{45}}{M_{45}}{j_{k_6}}{m_{k_6}}{J}{M} \notag \\ &\times
\tilde t^{60(2)}_{k_1k_2 k_3 k_4 k_5 k_6}\, ,
\end{align}
where in both bra and ket states the first two quasi-particle labels are coupled to intermediate two-body angular momenta $J_{12}$ and $J_{45}$, respectively. Whereas the angular-momentum coupling of the double BCC amplitude matrix elements is comparatively simple, performing it for the 20 terms defining the mode-6 tensor of present interest is much more challenging and error prone due to the increasing number of CGCs. Additionally, to the four CGCs arising from the external coupling there are four additional ones from inserting the transformations from~\eqref{eq:omegaJ}. Overall this requires the angular-momentum reduction of a string of eight CGCs. In order to deal with this problem systematically, a graph-theory based tool for automatized angular-momentum algebra of many-body tensor networks has been used~\cite{ripoche19b}. Much larger strings of CGCs can actually be treated in this framework, thus paving the way to relax state-of-the-art many-body truncations and extend high-accuracy \textit{ab initio} calculations of open-shell nuclei. For benchmarking purposes, an independent $m$-scheme code was implemented to verify the correctness of the results from the $J$-scheme implementation in small model spaces. 

Omitting the harmless energy denominator that only depends on external indices, the $J$-coupled form of the 20 contributions is given by
\begin{strip}
\begin{subequations}
\begin{align}
&\mathbf{T^{(3)}_{1}}: \sum_{n_{k_7} l_{k_7} t_{k_7} }\hat{J}^{-2}\hat{J_{12}}\hat{J_{45}}  {}^{J_{12}}\tilde{\Omega}^{31}_{\tilde{k}_{1}\tilde{k}_{2}\tilde{k}_{3}(n_{k_7} l_{k_7} t_{k_7} J) }{}^{J_{45}}\tilde t^{40(1)}_{\tilde{k}_{4}\tilde{k}_{5}\tilde{k}_{6}(n_{k_7} l_{k_7} t_{k_7} J)}  \\
&\mathbf{T^{(3)}_{2}}: \sum_{\tilde{k}_{7}J^{\prime\prime}}(-1)^{j_{{k}_{4}} +j_{{k}_{5}} +J_{45} }\hat{J_{12}}\hat{J_{45}}\hat{J^{\prime\prime}}^{2}  {}^{J_{12}}\tilde{\Omega}^{31}_{\tilde{k}_{1}\tilde{k}_{2}\tilde{k}_{5}\tilde{k}_{7}}{}^{J^{\prime\prime}}\tilde t^{40(1)}_{\tilde{k}_{3}\tilde{k}_{7}\tilde{k}_{6}\tilde{k}_{4}}  \sixj{j_{{k}_{3}}}{j_{{k}_{7}}}{J^{\prime\prime}}{j_{{k}_{5}}}{J}{J_{12}}\sixj{j_{{k}_{6}}}{j_{{k}_{4}}}{J^{\prime\prime}}{j_{{k}_{5}}}{J}{J_{45}} \\
&\mathbf{T^{(3)}_{3}}:  \sum_{\tilde{k}_{7}J^{\prime}J^{\prime\prime}}(-1)^{j_{{k}_{1}} +j_{{k}_{2}} +J_{12} }\hat{J_{12}}\hat{J_{45}}\hat{J^{\prime}}^{2}\hat{J^{\prime\prime}}^{2}  {}^{J^{\prime}}\tilde{\Omega}^{31}_{\tilde{k}_{3}\tilde{k}_{1}\tilde{k}_{4}\tilde{k}_{7}}{}^{J^{\prime\prime}}\tilde t^{40(1)}_{\tilde{k}_{2}\tilde{k}_{7}\tilde{k}_{6}\tilde{k}_{5}}  \sixj{j_{{k}_{3}}}{j_{{k}_{1}}}{J^{\prime}}{j_{{k}_{2}}}{J}{J_{12}}\sixj{j_{{k}_{2}}}{j_{{k}_{7}}}{J^{\prime\prime}}{j_{{k}_{4}}}{J}{J^{\prime}}\sixj{j_{{k}_{6}}}{j_{{k}_{5}}}{J^{\prime\prime}}{j_{{k}_{4}}}{J}{J_{45}} \\
&\mathbf{T^{(3)}_{4}}:  \sum_{\tilde{k}_{7}J^{\prime}}(-1)^{j_{{k}_{1}} +j_{{k}_{2}} +J_{12} }\hat{J_{12}}\hat{J_{45}}\hat{J^{\prime}}^{2}  {}^{J^{\prime}}\tilde{\Omega}^{31}_{\tilde{k}_{3}\tilde{k}_{1}\tilde{k}_{6}\tilde{k}_{7}}{}^{J_{45}}\tilde t^{40(1)}_{\tilde{k}_{2}\tilde{k}_{7}\tilde{k}_{4}\tilde{k}_{5}}  \sixj{j_{{k}_{3}}}{j_{{k}_{1}}}{J^{\prime}}{j_{{k}_{2}}}{J}{J_{12}}\sixj{j_{{k}_{2}}}{j_{{k}_{7}}}{J_{45}}{j_{{k}_{6}}}{J}{J^{\prime}} \\
& \mathbf{T^{(3)}_{5}}: - \sum_{\tilde{k}_{7}J^{\prime\prime}}\hat{J_{12}}\hat{J_{45}}\hat{J^{\prime\prime}}^{2}  {}^{J_{45}}\tilde{\Omega}^{31}_{\tilde{k}_{4}\tilde{k}_{5}\tilde{k}_{1}\tilde{k}_{7}}{}^{J^{\prime\prime}}\tilde t^{40(1)}_{\tilde{k}_{3}\tilde{k}_{2}\tilde{k}_{6}\tilde{k}_{7}}  \sixj{j_{{k}_{3}}}{j_{{k}_{2}}}{J^{\prime\prime}}{j_{{k}_{1}}}{J}{J_{12}}\sixj{j_{{k}_{6}}}{j_{{k}_{7}}}{J^{\prime\prime}}{j_{{k}_{1}}}{J}{J_{45}} \\
&\mathbf{T^{(3)}_{6}}: - \sum_{\tilde{k}_{7}J^{\prime}J^{\prime\prime}}\hat{J_{12}}\hat{J_{45}}\hat{J^{\prime}}^{2}\hat{J^{\prime\prime}}^{2}  {}^{J^{\prime}}\tilde{\Omega}^{31}_{\tilde{k}_{6}\tilde{k}_{5}\tilde{k}_{1}\tilde{k}_{7}}{}^{J^{\prime\prime}}\tilde t^{40(1)}_{\tilde{k}_{3}\tilde{k}_{2}\tilde{k}_{4}\tilde{k}_{7}}  \sixj{j_{{k}_{3}}}{j_{{k}_{2}}}{J^{\prime\prime}}{j_{{k}_{1}}}{J}{J_{12}}\sixj{j_{{k}_{1}}}{j_{{k}_{7}}}{J^{\prime}}{j_{{k}_{4}}}{J}{J^{\prime\prime}}\sixj{j_{{k}_{6}}}{j_{{k}_{5}}}{J^{\prime}}{j_{{k}_{4}}}{J}{J_{45}} \\
&\mathbf{T^{(3)}_{7}}: \sum_{\tilde{k}_{7}J^{\prime}J^{\prime\prime}}(-1)^{j_{{k}_{4}} +j_{{k}_{5}} +J_{45} }\hat{J_{12}}\hat{J_{45}}\hat{J^{\prime}}^{2}\hat{J^{\prime\prime}}^{2}  {}^{J^{\prime}}\tilde{\Omega}^{31}_{\tilde{k}_{3}\tilde{k}_{2}\tilde{k}_{5}\tilde{k}_{7}}{}^{J^{\prime\prime}}\tilde t^{40(1)}_{\tilde{k}_{1}\tilde{k}_{7}\tilde{k}_{6}\tilde{k}_{4}}  \sixj{j_{{k}_{3}}}{j_{{k}_{2}}}{J^{\prime}}{j_{{k}_{1}}}{J}{J_{12}}\sixj{j_{{k}_{1}}}{j_{{k}_{7}}}{J^{\prime\prime}}{j_{{k}_{5}}}{J}{J^{\prime}}\sixj{j_{{k}_{6}}}{j_{{k}_{4}}}{J^{\prime\prime}}{j_{{k}_{5}}}{J}{J_{45}} \\
&\mathbf{T^{(3)}_{8}}:-\sum_{\tilde{k}_{7}J^{\prime}J^{\prime\prime}}(-1)^{j_{{k}_{1}} +j_{{k}_{2}} +j_{{k}_{4}} +j_{{k}_{5}} +J_{12} +J_{45} }\hat{J_{12}}\hat{J_{45}}\hat{J^{\prime}}^{2}\hat{J^{\prime\prime}}^{2}  {}^{J^{\prime}}\tilde{\Omega}^{31}_{\tilde{k}_{6}\tilde{k}_{4}\tilde{k}_{2}\tilde{k}_{7}}{}^{J^{\prime\prime}}\tilde t^{40(1)}_{\tilde{k}_{3}\tilde{k}_{1}\tilde{k}_{5}\tilde{k}_{7}} \notag \\ & \hspace{2cm} \times \sixj{j_{{k}_{3}}}{j_{{k}_{1}}}{J^{\prime\prime}}{j_{{k}_{2}}}{J}{J_{12}} \sixj{j_{{k}_{2}}}{j_{{k}_{7}}}{J^{\prime}}{j_{{k}_{5}}}{J}{J^{\prime\prime}}\sixj{j_{{k}_{6}}}{j_{{k}_{4}}}{J^{\prime}}{j_{{k}_{5}}}{J}{J_{45}} \\
&\mathbf{T^{(3)}_{9}}: -\sum_{\tilde{k}_{7}}\hat{J_{12}}\hat{J_{45}}  {}^{J_{45}}\tilde{\Omega}^{31}_{\tilde{k}_{4}\tilde{k}_{5}\tilde{k}_{3}\tilde{k}_{7}}{}^{J_{12}}\tilde t^{40(1)}_{\tilde{k}_{1}\tilde{k}_{2}\tilde{k}_{6}\tilde{k}_{7}}  \sixj{j_{{k}_{6}}}{j_{{k}_{7}}}{J_{12}}{j_{{k}_{3}}}{J}{J_{45}} \\
&\mathbf{T^{(3)}_{10}}: -\sum_{\tilde{k}_{7}J^{\prime}}\hat{J_{12}}\hat{J_{45}}\hat{J^{\prime}}^{2}  {}^{J^{\prime}}\tilde{\Omega}^{31}_{\tilde{k}_{6}\tilde{k}_{5}\tilde{k}_{3}\tilde{k}_{7}}{}^{J_{12}}\tilde t^{40(1)}_{\tilde{k}_{1}\tilde{k}_{2}\tilde{k}_{4}\tilde{k}_{7}}  \sixj{j_{{k}_{3}}}{j_{{k}_{7}}}{J^{\prime}}{j_{{k}_{4}}}{J}{J_{12}}\sixj{j_{{k}_{6}}}{j_{{k}_{5}}}{J^{\prime}}{j_{{k}_{4}}}{J}{J_{45}} \\
&\mathbf{T^{(3)}_{11}}:  -\sum_{\tilde{k}_{7}J^{\prime\prime}}\hat{J_{12}}\hat{J_{45}}\hat{J^{\prime\prime}}^{2}  {}^{J_{12}}\tilde{\Omega}^{31}_{\tilde{k}_{1}\tilde{k}_{2}\tilde{k}_{4}\tilde{k}_{7}}{}^{J^{\prime\prime}}\tilde t^{40(1)}_{\tilde{k}_{3}\tilde{k}_{7}\tilde{k}_{6}\tilde{k}_{5}}  \sixj{j_{{k}_{3}}}{j_{{k}_{7}}}{J^{\prime\prime}}{j_{{k}_{4}}}{J}{J_{12}}\sixj{j_{{k}_{6}}}{j_{{k}_{5}}}{J^{\prime\prime}}{j_{{k}_{4}}}{J}{J_{45}} \\
&\mathbf{T^{(3)}_{12}}: -\sum_{\tilde{k}_{7}}\hat{J_{12}}\hat{J_{45}}  {}^{J_{12}}\tilde{\Omega}^{31}_{\tilde{k}_{1}\tilde{k}_{2}\tilde{k}_{6}\tilde{k}_{7}}{}^{J_{45}}\tilde t^{40(1)}_{\tilde{k}_{3}\tilde{k}_{7}\tilde{k}_{4}\tilde{k}_{5}}  \sixj{j_{{k}_{3}}}{j_{{k}_{7}}}{J_{45}}{j_{{k}_{6}}}{J}{J_{12}} \\
&\mathbf{T^{(3)}_{13}}: -\sum_{\tilde{k}_{7}J^{\prime}J^{\prime\prime}}(-1)^{j_{{k}_{1}} +j_{{k}_{2}} +j_{{k}_{4}} +j_{{k}_{5}} +J_{12} +J_{45} }\hat{J_{12}}\hat{J_{45}}\hat{J^{\prime}}^{2}\hat{J^{\prime\prime}}^{2}  {}^{J^{\prime}}\tilde{\Omega}^{31}_{\tilde{k}_{3}\tilde{k}_{1}\tilde{k}_{5}\tilde{k}_{7}}{}^{J^{\prime\prime}}\tilde t^{40(1)}_{\tilde{k}_{2}\tilde{k}_{7}\tilde{k}_{6}\tilde{k}_{4}} \notag \\ & \hspace{2cm} \times \sixj{j_{{k}_{3}}}{j_{{k}_{1}}}{J^{\prime}}{j_{{k}_{2}}}{J}{J_{12}}\sixj{j_{{k}_{2}}}{j_{{k}_{7}}}{J^{\prime\prime}}{j_{{k}_{5}}}{J}{J^{\prime}}\sixj{j_{{k}_{6}}}{j_{{k}_{4}}}{J^{\prime\prime}}{j_{{k}_{5}}}{J}{J_{45}} \\
&\mathbf{T^{(3)}_{14}}:  \sum_{\tilde{k}_{7}J^{\prime}J^{\prime\prime}}(-1)^{j_{{k}_{4}} +j_{{k}_{5}} +J_{45} }\hat{J_{12}}\hat{J_{45}}\hat{J^{\prime}}^{2}\hat{J^{\prime\prime}}^{2}  {}^{J^{\prime}}\tilde{\Omega}^{31}_{\tilde{k}_{6}\tilde{k}_{4}\tilde{k}_{1}\tilde{k}_{7}}{}^{J^{\prime\prime}}\tilde t^{40(1)}_{\tilde{k}_{3}\tilde{k}_{2}\tilde{k}_{5}\tilde{k}_{7}}  \sixj{j_{{k}_{3}}}{j_{{k}_{2}}}{J^{\prime\prime}}{j_{{k}_{1}}}{J}{J_{12}}\sixj{j_{{k}_{1}}}{j_{{k}_{7}}}{J^{\prime}}{j_{{k}_{5}}}{J}{J^{\prime\prime}}\sixj{j_{{k}_{6}}}{j_{{k}_{4}}}{J^{\prime}}{j_{{k}_{5}}}{J}{J_{45}} \\
&\mathbf{T^{(3)}_{15}}: - \sum_{\tilde{k}_{7}J^{\prime}J^{\prime\prime}}\hat{J_{12}}\hat{J_{45}}\hat{J^{\prime}}^{2}\hat{J^{\prime\prime}}^{2}  {}^{J^{\prime}}\tilde{\Omega}^{31}_{\tilde{k}_{3}\tilde{k}_{2}\tilde{k}_{4}\tilde{k}_{7}}{}^{J^{\prime\prime}}\tilde t^{40(1)}_{\tilde{k}_{1}\tilde{k}_{7}\tilde{k}_{6}\tilde{k}_{5}}  \sixj{j_{{k}_{3}}}{j_{{k}_{2}}}{J^{\prime}}{j_{{k}_{1}}}{J}{J_{12}}\sixj{j_{{k}_{1}}}{j_{{k}_{7}}}{J^{\prime\prime}}{j_{{k}_{4}}}{J}{J^{\prime}}\sixj{j_{{k}_{6}}}{j_{{k}_{5}}}{J^{\prime\prime}}{j_{{k}_{4}}}{J}{J_{45}} \\ 
&\mathbf{T^{(3)}_{16}}: - \sum_{\tilde{k}_{7}J^{\prime}}\hat{J_{12}}\hat{J_{45}}\hat{J^{\prime}}^{2}  {}^{J^{\prime}}\tilde{\Omega}^{31}_{\tilde{k}_{3}\tilde{k}_{2}\tilde{k}_{6}\tilde{k}_{7}}{}^{J_{45}}\tilde t^{40(1)}_{\tilde{k}_{1}\tilde{k}_{7}\tilde{k}_{4}\tilde{k}_{5}}  \sixj{j_{{k}_{3}}}{j_{{k}_{2}}}{J^{\prime}}{j_{{k}_{1}}}{J}{J_{12}}\sixj{j_{{k}_{1}}}{j_{{k}_{7}}}{J_{45}}{j_{{k}_{6}}}{J}{J^{\prime}}\\
&\mathbf{T^{(3)}_{17}}:  \sum_{\tilde{k}_{7}J^{\prime\prime}}(-1)^{j_{{k}_{1}} +j_{{k}_{2}} +J_{12} }\hat{J_{12}}\hat{J_{45}}\hat{J^{\prime\prime}}^{2}  {}^{J_{45}}\tilde{\Omega}^{31}_{\tilde{k}_{4}\tilde{k}_{5}\tilde{k}_{2}\tilde{k}_{7}}{}^{J^{\prime\prime}}\tilde t^{40(1)}_{\tilde{k}_{3}\tilde{k}_{1}\tilde{k}_{6}\tilde{k}_{7}}  \sixj{j_{{k}_{3}}}{j_{{k}_{1}}}{J^{\prime\prime}}{j_{{k}_{2}}}{J}{J_{12}}\sixj{j_{{k}_{6}}}{j_{{k}_{7}}}{J^{\prime\prime}}{j_{{k}_{2}}}{J}{J_{45}} \\
&\mathbf{T^{(3)}_{18}}:  \sum_{\tilde{k}_{7}J^{\prime}J^{\prime\prime}}(-1)^{j_{{k}_{1}} +j_{{k}_{2}} +J_{12} }\hat{J_{12}}\hat{J_{45}}\hat{J^{\prime}}^{2}\hat{J^{\prime\prime}}^{2}  {}^{J^{\prime}}\tilde{\Omega}^{31}_{\tilde{k}_{6}\tilde{k}_{5}\tilde{k}_{2}\tilde{k}_{7}}{}^{J^{\prime\prime}}\tilde t^{40(1)}_{\tilde{k}_{3}\tilde{k}_{1}\tilde{k}_{4}\tilde{k}_{7}}  \sixj{j_{{k}_{3}}}{j_{{k}_{1}}}{J^{\prime\prime}}{j_{{k}_{2}}}{J}{J_{12}}\sixj{j_{{k}_{2}}}{j_{{k}_{7}}}{J^{\prime}}{j_{{k}_{4}}}{J}{J^{\prime\prime}}\sixj{j_{{k}_{6}}}{j_{{k}_{5}}}{J^{\prime}}{j_{{k}_{4}}}{J}{J_{45}} \\
&\mathbf{T^{(3)}_{19}}:  \sum_{\tilde{k}_{7}J^{\prime}}(-1)^{j_{{k}_{4}} +j_{{k}_{5}} +J_{45} }\hat{J_{12}}\hat{J_{45}}\hat{J^{\prime}}^{2}  {}^{J^{\prime}}\tilde{\Omega}^{31}_{\tilde{k}_{6}\tilde{k}_{4}\tilde{k}_{3}\tilde{k}_{7}}{}^{J_{12}}\tilde t^{40(1)}_{\tilde{k}_{1}\tilde{k}_{2}\tilde{k}_{5}\tilde{k}_{7}}  \sixj{j_{{k}_{3}}}{j_{{k}_{7}}}{J^{\prime}}{j_{{k}_{5}}}{J}{J_{12}}\sixj{j_{{k}_{6}}}{j_{{k}_{4}}}{J^{\prime}}{j_{{k}_{5}}}{J}{J_{45}} \\
&\mathbf{T^{(3)}_{20}}: \sum_{n_{k_7} l_{k_7} t_{k_7} }\hat{J}^{-2}\hat{J_{12}}\hat{J_{45}}  {}^{J_{45}}\tilde{\Omega}^{31}_{\tilde{k}_{4}\tilde{k}_{5}\tilde{k}_{6} (n_{k_7} l_{k_7} t_{k_7} J)}{}^{J_{12}}\tilde t^{40(1)}_{\tilde{k}_{1}\tilde{k}_{2}\tilde{k}_{3} (n_{k_7} l_{k_7} t_{k_7} J)}  
\end{align}
\label{eq:kappa2J}
\end{subequations}
To the second-order triple amplitude is associated the fourth-order perturbative ground-state energy correction
\begin{align}
\Delta \Omega^{[4_T]}_0 = \sum_{ \substack{k_1 k_2 k_3 \\ k_4 k_5 k_6} }  |t^{60 (2)}_{k_1 k_2 k_3 k_4 k_5 k_6} |^2  E_{k_1 k_2 k_3 k_4 k_5 k_6} \, ,
\label{eq:Etriples}
\end{align}
which is a $N^6$ process. The $J$-coupled form of~\eqref{eq:Etriples} is given by
\begin{align}
\Delta \Omega^{[4_T]}_0 = \sum_{J_{12} J_{45} J} \sum_{\substack{\tilde k_1 \tilde k_2 \tilde k_3 \\ \tilde k_4 \tilde k_5 \tilde k_6 }} (2 J +1 ) |{^{J_{12} J_{45} J}} \tilde t^{60(2)}_{\tilde k_1 \tilde k_2 \tilde k_3 \tilde k_4 \tilde k_5 \tilde k_6}|^2 E_{\tilde k_1 \tilde k_2 \tilde k_3 \tilde k_4  \tilde k_5 \tilde k_6}\, ,
\label{eq:EtriplesJ}
\end{align}
and requires $J^2_{2\text{max}} J_{3\text{max}} \cdot \tilde N^6$ evaluations, where $J_{k\text{max}}$ denotes the number of $k$-body angular momentum channels. For the case $k=3$ this number of channels is given by $J_{3\text{max}}=3e_\text{max} +2$ as long as no additional truncation on $e_{3\text{max}} \equiv e_1 + e_2 + e_3$ is employed.
\end{strip}

Due to working in quasi-particle basis, the evaluation of Eq.~\eqref{eq:kappa2J} and Eq.~\eqref{eq:EtriplesJ} is a computationally challenging task requiring orders of magnitudes more resources than its symmetry-conserving counterpart in closed-shell nuclei. Taking mid-mass Ca or Ni isotopes as an example, the runtime increases by a factor $10^2-10^3$ assuming an $e_\text{max}=12$ model space corresponding to $\tilde N = 182$.

Finally, it is worth noting that a consistent inclusion of \textit{all permutations} is mandatory. From~\eqref{eq:kappa2J} it is obvious that the evaluation of some of the terms is computationally simpler than others due to the appearance of additional recoupling symbols accompanied by angular-momentum summations. However, from an $m$-scheme perspective all terms are equally important and the difference in computational complexity is just an artefact of the chosen external coupling order. In actual applications, different contributions corresponding to recoupling patterns of different complexity (i.e. different number of $6j$-symbols) are equally important such that a consistent inclusion of all terms is necessary.

\end{appendix}

\bibliographystyle{spphys}
\bibliography{bib_nucl}

\end{document}